\documentclass[10pt,journal,compsoc]{IEEEtran}
\usepackage{mathtools}
\usepackage{amsfonts}
\usepackage{booktabs}
\usepackage{algorithm}
\usepackage{algpseudocode}
\usepackage{float}
\usepackage{hyperref}
\usepackage{amssymb}

\ifCLASSOPTIONcompsoc
  \usepackage[nocompress]{cite}
\else
  \usepackage{cite}
\fi
\ifCLASSINFOpdf
\else
\fi
\begin{document}
%
\title{\huge Universal Source Separation with Weakly Labelled Data}

\author{Qiuqiang Kong*, Ke Chen*, Haohe Liu, Xingjian Du \\ Taylor Berg-Kirkpatrick, Shlomo Dubnov, Mark D. Plumbley

\IEEEcompsocitemizethanks{\IEEEcompsocthanksitem Qiuqiang Kong and Ke Chen contribute equally to this work. Qiuqiang Kong: kongqiuqiang@bytedance.com, Xingjian Du: duxingjian.real@bytedance.com are with ByteDance, Shanghai, China. Ke Chen: knutchen@ucsd.edu, Shlomo Dubnov: sdubnov@ucsd.edu, and Taylor Berg-Kirkpatrick: tberg@eng.ucsd.edu are with University of California San Diego, San Diego, USA. Haohe Liu: haohe.liu@surrey.ac.uk, Mark D. Plumbley: m.plumbley@surrey.ac.uk are with University of Surrey, Guildford, UK. 

}
}

%
%

\markboth{Journal of \LaTeX\ Class Files,~Vol.~14, No.~8, August~2015}%
{Shell \MakeLowercase{\textit{et al.}}: Bare Demo of IEEEtran.cls for Computer Society Journals}
%



\IEEEtitleabstractindextext{%
\begin{abstract}

Universal source separation (USS) is a fundamental research task for computational auditory scene analysis, which aims to separate mono recordings into individual source tracks. There are three potential challenges awaiting the solution to the audio source separation task. First, previous audio source separation systems mainly focus on separating one or a limited number of specific sources. There is a lack of research on building a unified system that can separate arbitrary sources via a single model. Second, most previous systems require clean source data to train a separator, while clean source data are scarce. Third, there is a lack of USS system that can automatically detect and separate active sound classes in a hierarchical level. To use large-scale weakly labeled/unlabeled audio data for audio source separation, we propose a universal audio source separation framework containing: 1) an audio tagging model trained on weakly labeled data as a query net; and 2) a conditional source separation model that takes query net outputs as conditions to separate arbitrary sound sources. We investigate various query nets, source separation models, and training strategies and propose a hierarchical USS strategy to automatically detect and separate sound classes from the AudioSet ontology. By solely leveraging the weakly labelled AudioSet, our USS system is successful in separating a wide variety of sound classes, including sound event separation, music source separation, and speech enhancement. The USS system achieves an average signal-to-distortion ratio improvement (SDRi) of 5.57 dB over 527 sound classes of AudioSet; 10.57 dB on the DCASE 2018 Task 2 dataset; 8.12 dB on the MUSDB18 dataset; an SDRi of 7.28 dB on the Slakh2100 dataset; and an SSNR of 9.00 dB on the voicebank-demand dataset. We release the source code at \url{https://github.com/bytedance/uss}

\end{abstract}

}

\maketitle

\IEEEdisplaynontitleabstractindextext

%
\IEEEpeerreviewmaketitle

\IEEEraisesectionheading{\section{Introduction}\label{sec:introduction}}

\IEEEPARstart{M}{ono} source separation is the task of separating single-channel audio recordings into individual source tracks. An audio recording may consist of several sound events and acoustic scenes. 
\textit{Universal source separation} (USS) is a task to separate arbitrary sound from a recording. Source separation has been researched for several years and has a wide range of applications, including speech enhancement \cite{loizou2007speech, xu2014regression}, music source separation \cite{stoter20182018}, and sound event separation\cite{heittola2011sound, turpault2020improving}. USS is closely related to the well-known cocktail party problem \cite{haykin2005cocktail}, where sounds from different sources in the world mix in the air before arriving at the ear, requiring the brain to estimate individual sources from the received mixture. Humans can focus on a particular sound source and separate it from others, a skill sometimes called \textit{selective hearing}. As a study of auditory scene analysis by computational means, computational auditory scene analysis \cite{brown1994computational, wang2006computational} systems are machine listening systems that aim to separate mixtures of sound sources in the same way that human listeners do. 

Many previous works mainly focus on \textit{specific source separation} that only separate one or a few sources; these include speech separation \cite{loizou2007speech, xu2014regression} and music source separation \cite{stoller2018wave} tasks. Different from specific source separation tasks such as speech enhancement or music source separation, a USS system aims to automatically detect and separate the tracks of sound sources from a mixture. One difficulty of USS is that there are hundreds of different sounds in the world, and it is difficult to separate all sounds using a unified model \cite{luo2021rethinking}. Recently, the USS problem has attracted the interests of several researchers. A system \cite{kavalerov2019universal} was proposed to separate arbitrary sounds by predicting the masks of sounds, where the masks control how many signals should remain from the mixture signals. Unsupervised USS systems \cite{wisdom2020unsupervised, tzinis2020sudo} were proposed to separate sounds by mixing training samples into a mixture and separating the mixture into a variable number of sources. A Free Universal Sound Separation (FUSS) system applied a time-domain convolutional network (TDCN++) system to separate mixtures into up to 4 separate sources. A class conditions system \cite{seetharaman2019class} was proposed for 4-stem music source separation. Some other methods \cite{tzinis2020improving, wang2018voicefilter, gfeller2021one} use audio embeddings that character an audio clip to control what sources to separate from a mixture. In \cite{choi2021lasaft}, one-hot encodings of sound classes are used as controls to separate corresponding sources. Other sound separation systems include learning to separate from weakly labelled scenes \cite{pishdadian2020finding} and SuDoRM-RM \cite{tzinis2020sudo, tzinis2021compute} to remove sound sources from mixtures. Recently, a language-based source separation system was proposed in \cite{liu2022separate}. 
Those systems are mainly trained on small datasets and do not scale to automatically detect and separate hundreds of sound classes.

\begin{figure*}[t]
    \centering
    \includegraphics[width=\textwidth]{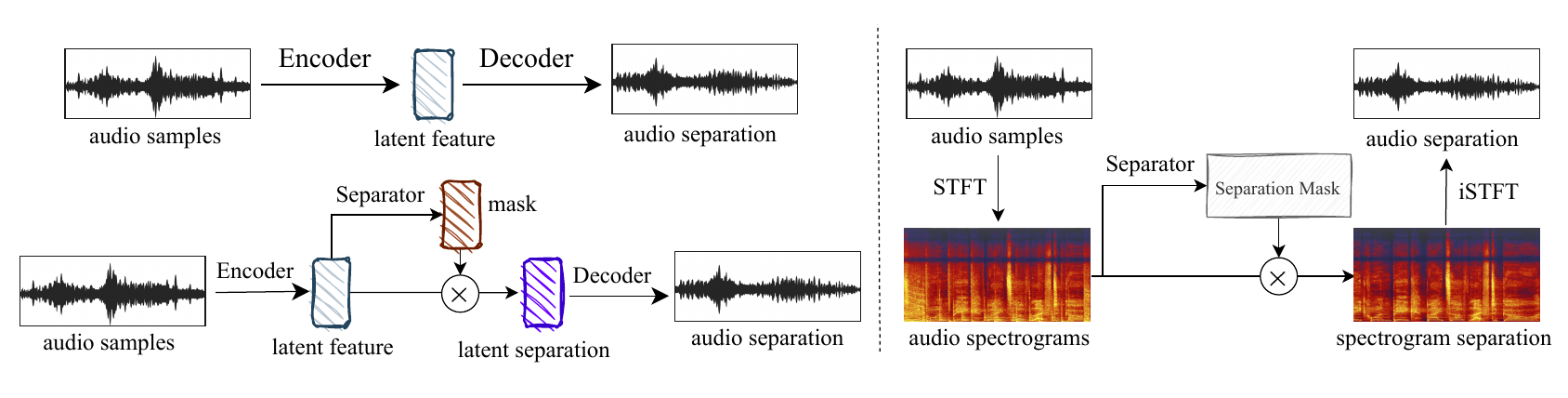}
    \caption{The standard architecture of deep-learning-based audio source separation model. Left top: synthesis-based separation model. Left bottom: mask-based separation model. Right: the general type of frequency-domain separation model.}
    \label{fig:ssp-arch}
\end{figure*}

For the source separation problem, we define \textit{clean source data} as audio segments that contain only target sources without other sources. The clean source data can be mixed to form audio mixtures to train potential separators. 
 However, collecting clean data is time-consuming. For sound classes such as ``Speech'', clean data can be recorded in the laboratory. But for many other environmental sounds such as ``Thunder'', the collection of clean data is difficult. Recently, the concept of \textit{weakly labelled data} \cite{mesaros2019sound, gemmeke2017audio} was used in audio signal processing. In contrast to clean source data, weakly labelled data contains multiple sources in an audio clip. An audio clip is labelled with one or multiple tags, while the time information of tags is unknown. For example, a 10-second audio clip is labelled as  ``Thunder'' and ``Rain'', but the time when these two events exactly appear within this 10-second clip is not provided. Weakly labelled data has been widely used in audio tagging \cite{shah2018closer, kong2020panns, gong2021psla, chen2022hts} and sound event detection \cite{kong2017joint, wang2018polyphonic, adavanne2019sound}. 
 But there has been limited work on using weakly labelled for source separation \cite{kong2020source, chen2022zero}.

In this work, we propose a USS framework that can be trained with weakly labelled data. This work extends our previously proposed USS systems \cite{kong2020source, kong2021decoupling, chen2022zero} with contributions as follows:
\begin{itemize}
    \item We are the first to use large-scale weakly labelled data to train USS systems that can separate hundreds of sound classes.
    \item We propose to use sound event detection systems trained on weakly labelled data to detect short segments that are most likely to contain sound events. 
    \item We investigate a variety of query nets to extract conditions to build USS systems. The query nets are pretrained or finetuned audio tagging systems.
    \item We propose a hierarchical USS strategy to automatically detect and separate the sources of existing sound classes with an hierarchical AudioSet ontology. The USS procedure do require the specification of the sound classes to separate.
    \item We show that a single USS system is able to perform a wide range of separation tasks, including sound event separation, music source separation, and speech enhancement. We conduct comprehensive ablation studies to investigate how different factors in our system affect the separation performance. 
    
\end{itemize}

This article is organized as follows. Section 2 introduces neural network-based source separation systems. Section 3 introduces our proposed weakly labelled source separation framework. Section 4 reports on the results of experiments. Section 5 concludes this work. 


\section{Source Separation via Neural Networks}\label{section:ss_strong}
Deep learning methods for audio source separation have outperformed traditional methods such as Non-negative Matrix Factorization \cite{nmf}. Fig. \ref{fig:ssp-arch} shows source separation models in the time domain (left) and in the frequency domain (right). Here, we introduce the basic methodology of those separation models. 
\subsection{Time-domain Separation Models}
A neural network time-domain separation model $f$ is typically constructed as an encoder-decoder architecture, as shown in the left of Fig. \ref{fig:ssp-arch}. Formally, given a single-channel audio clip $x \in \mathbb{R}^{L}$ and a separation target $s \in \mathbb{R}^{L}$, where $L$ is sample length, the separator $f$ contains two types: a synthesis-based separation system that directly outputs the waveform of the target source, and a mask-based separation that predict a mask that can be multiplied to the mixture to output the target source.

Separation models such as Demucs \cite{defossez2019demucs, defossez2019music} and Wave-U-Net \cite{stoller2018wave}, $f$ directly estimates the final separation target: $\hat{s} = f(x)$. 
Mask-based separation models such as TasNet \cite{luo2018tasnet} and ConvTasNet \cite{luo2019conv} predict masks in the latent space produced by the neural network. The masks control how much of sources should remain from the mixture. Then, a decoder is designed to reconstruct the separated waveform from the masked latent feature produced by the neural network.


\subsection{Frequency-domain Separation Models}
In contrast to time-domain models, frequency-domain models leverage a spectrogram, such as a short-time Fourier transform (STFT), to facilitate the separation process. Harmonic features have more patterns in the frequency domain than those in the time domain. This might help improve separation performance in source separation tasks, such as music source separation and environmental sound separation \cite{luo2022music}.

Formally, given a mono audio clip $x$, we denote the STFT of $x$ as a complex matrix $X \in \mathbb{C}^{T \times F}$, where $T$ is the number of time frames and $F$ is the number of frequency bins. We denote the magnitude and the phase of $ X $ as $|X|$ and $\angle X$, respectively. The right part of Fig. \ref{fig:ssp-arch} shows a frequency-domain separation system $f$ predicting a magnitude ideal ratio mask (IRM) \cite{narayanan2013ideal} $M \in \mathbb{R}^{T \times F}$ or a complex IRM (cIRM) \cite{williamson2015complex} $M \in \mathbb{C}^{T \times F}$ that can be multiplied by the STFT of the mixture to obtain the STFT of the separated source. The complex STFT of the separated source $ \hat{S} \in \mathbb{C}^{T \times F} $ can be calculated by:
\begin{align}
\hat{S}=M \odot X.
\end{align}
\noindent where $\odot$ is the element-wise complex multiplication. Then, the separated source $ \hat{s} \in \mathbb{R}^{L} $ can be obtained by applying an inverse STFT on $ \hat{S} $.

Frequency domain models include fully connected neural networks \cite{xu2014regression}, recurrent neural networks (RNNs) \cite{huang2015joint, takahashi2018mmdenselstm, uhlich2017improving}, and convolutional neural networks (CNNs) \cite{chandna2017monoaural, takahashi2018mmdenselstm, hu2020dccrn}. UNets \cite{jansson2017singing, hennequin2020spleeter} are variants of CNN that contain encoder and decoder layers for source separation. Band-split RNNs (BSRNNs) \cite{luo2022music} apply RNNs along both the time and frequency axes to capture time and frequency domain dependencies. There are also approaches such as hybrid Demucs \cite{defossez2021hybrid, rouard2022hybrid} which combine time and frequency domain systems to build source separation systems.


\subsection{Challenges of Source Separation Models}

As mentioned above, many previous source separation systems require clean source data to train source separation systems. However, the collection of clean source data is difficult and time-consuming. Table \ref{table:datasets} summarizes datasets that can be used for source separation. On the one hand, previous clean source datasets have durations of around tens of hours. On the other hand, weakly labelled datasets are usually larger than clean source datasets and clean datasets. AudioSet \cite{gemmeke2017audio} is a representative weakly labelled dataset containing over 5,800 hours of 10-second audio clips and is larger in both size and number of sound classes than clean source datasets. AudioSet has an ontology of 527 sound classes in its released version. The ontology of AudioSet has a tree structure, where each audio clip may contain multiple tags.

In this work, we use the weakly labelled AudioSet dataset containing 5,800 hours to train a USS system that can separate hundreds of sound classes.

\begin{table}[t]
\centering
\caption{Source separation datasets. The types can be clean data or weakly labelled data.}
\label{table:datasets}
\begin{tabular}{lccc}
 \toprule
 Dataset & Dur. (h) & Classes & Types \\
 \midrule
 Voicebank-Demand \cite{veaux2013voice} &  19 & 1 & Clean \\
 MUSDB18 \cite{rafii2017musdb18} &  6 & 4 & Clean \\
 UrbanSound8K \cite{salamon2014dataset} &  10 & 10 & Clean \\
 FSDKaggle 2018 \cite{fonseca2018general} & 20 & 41 & Clean \\
 FUSS \cite{wisdom2021s} & 23 & 357 & Clean \\
 AudioSet \cite{gemmeke2017audio} & 5,800 & 527 & Weak \\
 \bottomrule
\end{tabular}
\end{table}

\begin{figure}[t]
  \centering
  \centerline{\includegraphics[width=\columnwidth]{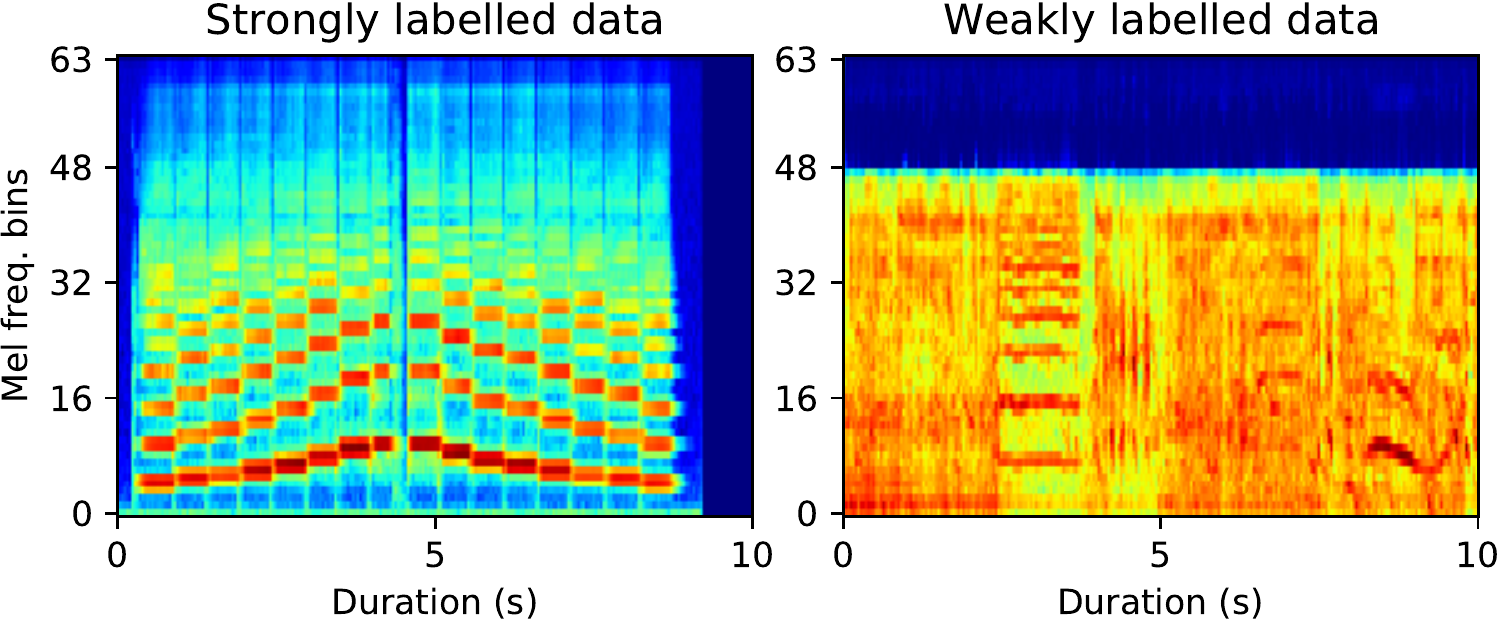}}
  \caption{Left: Clean source data of sound class ``Flute''. Right: Weakly labelled data of sound class ``Air horn, truck horn'' which only occurs between 2.5s - 4.0s.}
  \label{fig:strongweak}
\end{figure}


\section{USS with Weakly Labelled Data} \label{sec:sswithwld}

\subsection{Weakly Labelled Data}
In contrast to clean source data, weakly labelled data only contain the labels of what sound classes are present in an audio recording. Weakly labelled data may also contain interfering sounds. There are no time stamps for sound classes or clean sources. We denote the $n$-th audio clip in a weakly labelled dataset as $ a_{n} $ where $ a $ is the abbreviation for the \textit{audio}. The tags of $ a_{n} $ is denoted as $ y_{n} \in \{0, 1\}^{K} $, where $ K $ is the number of sound classes. The value $ y_{n}(k) = 1 $ indicates the presence of a sound class $ k $ while $ y_{n}(k) = 0 $ indicates the absence of a sound class $k$. We denote a weakly labelled dataset as $ D = \{a_{n}, y_{n}\}_{n=1}^{N} $, where $ N $ is the number of training samples in the dataset. The left part of Fig. \ref{fig:strongweak} shows a clean source audio clip containing the clean waveform of ``Flute''. The right part of Fig. \ref{fig:strongweak} shows a weakly labelled audio clip containing a target sound class ``Air horn, truck horn'' which only occurs between 2.5 s and 4.0 s. The weakly labelled audio recording also contains unknown interference sounds, i.e., $y_{n}(k)=0$ may contain missing tags for some sound class $k$.



\begin{figure}[t]
  \centering
  \includegraphics[width=\columnwidth]{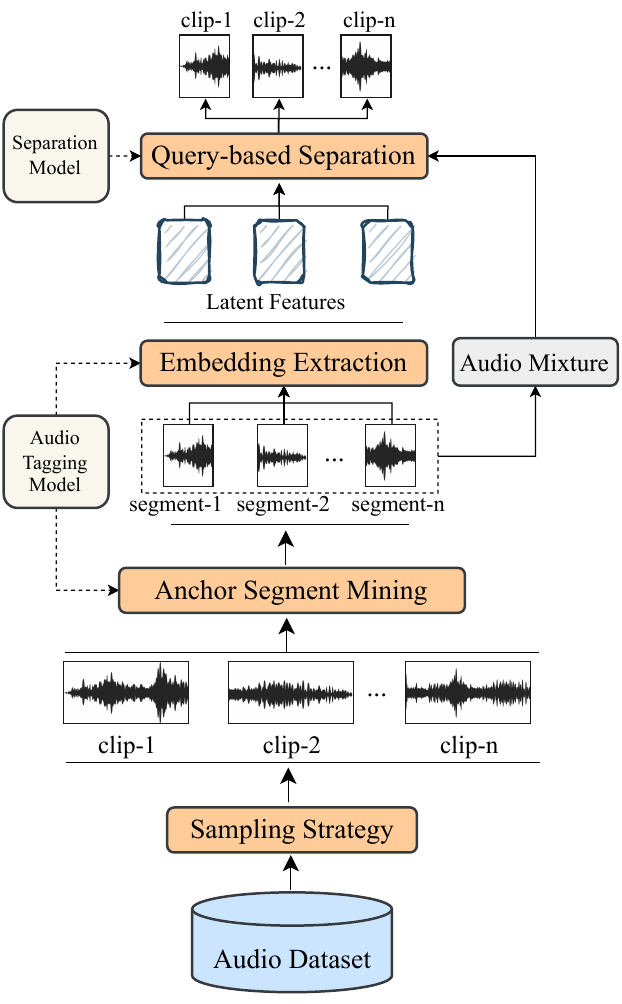}
  \caption{The architecture of our proposed query-based audio source separation pipeline trained from weakly-labeld data, including datasets, sampling strategies, audio tagging model, and conditional audio source separation models.}
  \label{fig:ssh_pipeline}
\end{figure}

The goal of a weakly labelled USS system is to separate arbitrary sounds trained with only weakly labelled data.
Fig. \ref{fig:ssh_pipeline} depicts the architecture of our proposed system, containing four steps: 

\begin{enumerate}
    \item We apply a sampling strategy to sample audio clips of different sound classes from a weakly labelled audio dataset. 
    \item We define an \textit{anchor segment} as a short segment that is most likely to contain a target sound class in a long audio clip. We apply an anchor segment mining algorithm to localize the occurrence of events/tags in the weakly labelled audio tracks.
    \item Use pretrained audio tagging models to predict the tag probabilities or embeddings of anchor segments.
    \item Mix anchor segments as input mixtures. Train a query-based separation network to separate the mixture into one of the target source queried by the sound class condition. 
\end{enumerate}

\subsection{Audio Clips Sampling}
For a large-scale dataset, we apply two sampling strategies: 1) random sampling: randomly sample audio clips from the dataset to constitute a mini-batch; and 2) balanced sampling: sample audio clips from different sound classes to constitute a mini-batch to ensure the clips contain different sound classes. AudioSet is highly unbalanced; sound classes such as ``Speech'' and ``Music'' have almost 1 million audio clips, while sound classes such as ``tooth breath'' have only tens of training samples. Without balanced sampling, the neural network may never see ``tooth breath'' if the training is not long enough. Following the training scheme of audio classification systems \cite{kong2020panns, gong2021ast,gong2021psla,chen2022hts}, we apply the balanced sampling strategy to retrieve audio data from AudioSet so that all sound classes can be sampled equally. That is, each sound class is sampled evenly from the unbalanced dataset. We denote a mini-batch of sampled audio clips as $ \{ a_{i} \}_{i=1}^{B} $, where $ B $ is the mini-batch size.

\begin{figure}[t]
  \centering
  \centerline{\includegraphics[width=\columnwidth]{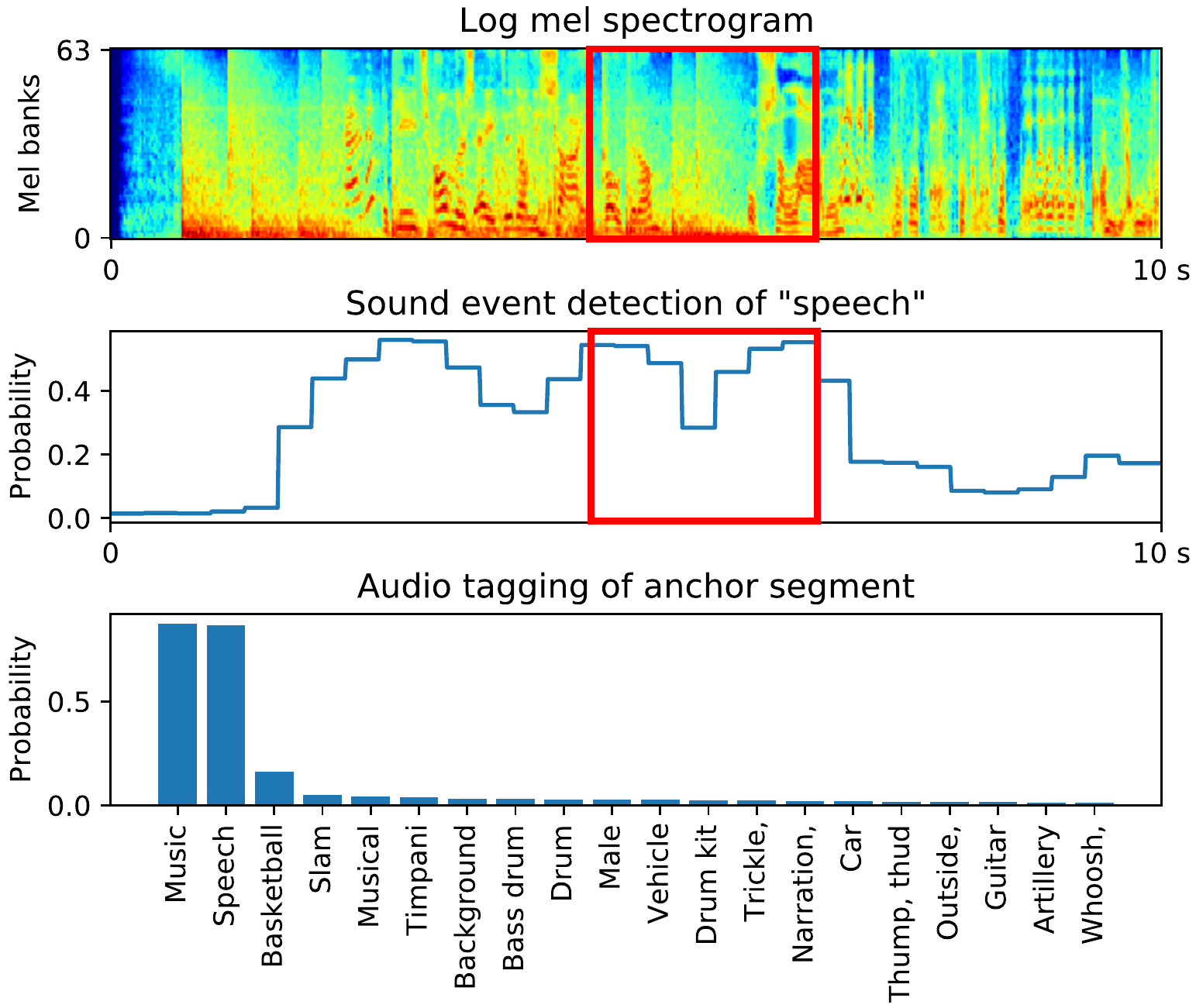}}
  \caption{Top: log mel spectrogram of a 10-second audio clip from AudioSet; Middle: predicted SED probability of “Speech”, where red block shows the selected anchor segment; Bottom: predicted audio tagging probabilities of the anchor segment.}
  \label{fig:10s_audioset_speech}
\end{figure}

\begin{figure*}[t]
    \centering
    \includegraphics[width=\textwidth]{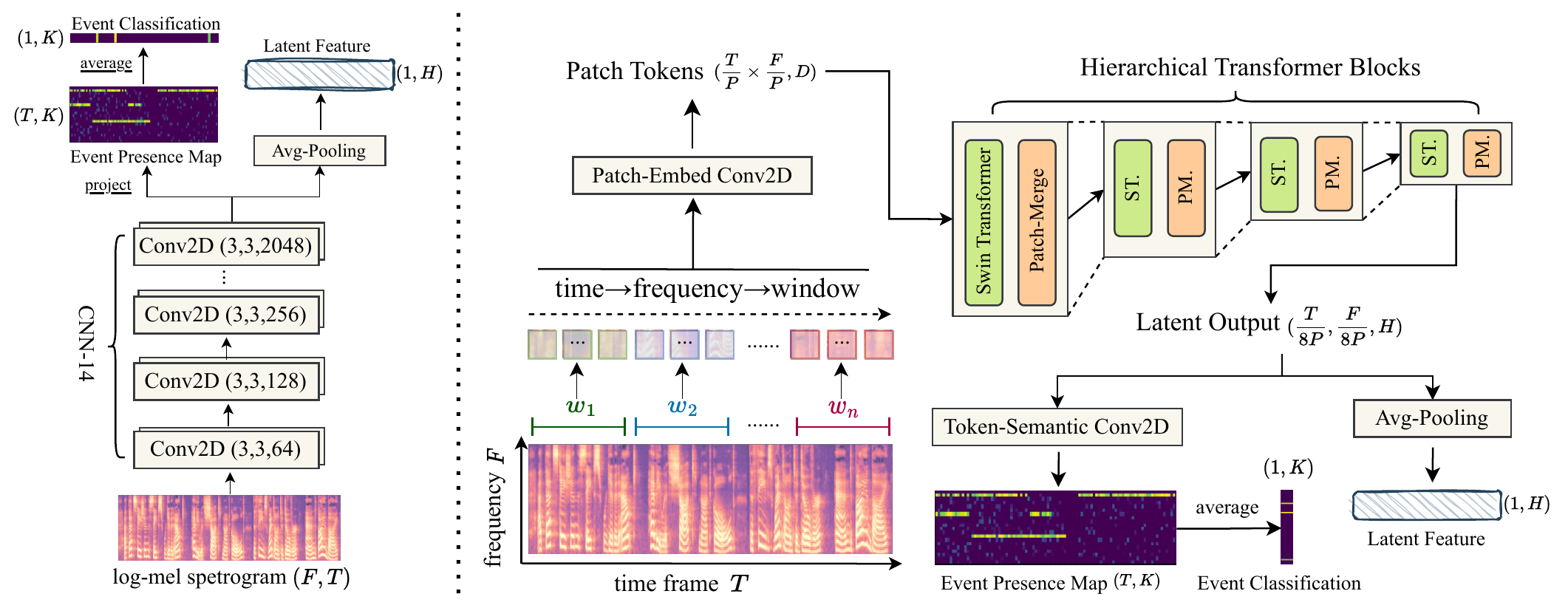}
    \caption{Two audio tagging models for audio classification, sound event detection, and latent feature production. Left: Pretrained Audio Neural Networks (PANN) in CNN14 architecture. Right: Hierarchical Token-Semantic Transformer (HTS-AT) in 4-block architecture.}
    \label{fig:pann-htsat}
\end{figure*}

\subsection{Anchor Segment Mining} \label{section:anchor}
We define \textit{anchor segment mining} as a procedure to localize anchor segments in an audio clip. 
We use sound event detection models that are trained only on weakly-labelled data but can localize the occurrence (i.e. time stamps) of sound classes. Recently, audio tagging systems trained with the weakly labelled AudioSet \cite{hershey2017cnn, kong2020panns, wang2019comparison, chen2022hts, gong2021psla} have outperformed systems trained with clean source data. We apply Pretrained Audio Neural Networks (PANNs) \cite{kong2020panns} and a Hierarchical Token-semantic Audio Transformer (HTS-AT) \cite{chen2022hts} as audio tagging models to perform the anchor segment mining procedure. Such models are able to extract audio clips with relatively clean sound sources from weakly labelled audio samples. 

Anchor segment mining is the core part of USS systems trained with weakly labelled data. Since the weakly labeled audio track does not always contain the labeled sound class throughout its timeline, we need to extract a short audio segment inside this track to create source data for training the separation model. Formally, given an audio clip $a_{i} \in \mathbb{R}^L$, an anchor segment mining algorithm extracts an anchor segment $s_{i} \in \mathbb{R}^{L'}$ from $a_{i}$, where $L' < L$ is the samples number of the anchor segment. 

For each audio clip in mini-batch $ \{ a_{i} \}_{i=1}^{B} $, we propose two types of anchor segment mining strategies: (A) Randomly select an anchor segment $ s_{i} $ from an audio clip $a_{i}$, or (B) Apply a pretrained sound event detection (SED) system to detect an anchor segment $ s_{i} $, where the center of $s_{i}$ is the time stamp where the sound class label is most likely to occur. For the SED method (B) of anchor mining, we leverage PANNs \cite{kong2020panns} and HTS-AT \cite{chen2022hts} to perform sound event detection on the audio track.

We introduce the SED anchor segment mining strategy (B) as follows. For an audio clip $ a_{i} $, a SED system produces two main outputs: 1) event classification prediction $p_{\text{AT}} \in [0, 1]^K$ where $K$ is the number of sound classes and $\text{AT}$ is the abbreviation for audio tagging; and 2) framewise event prediction $p_{\text{SED}} \in [0, 1]^{T \times K}$, where $T$ the number of frames and $\text{SED}$ is the abbreviation for sound event detection. To use weakly labelled data in training, the audio tagging prediction is usually calculated by maximizing the framewise event prediction $p_{\text{SED}}$. 
Both PANNs and HTS-AT are trained with the weakly labeled data by minimizing the binary cross entropy loss between the audio tagging prediction $p_{\text{AT}}$ and the labels $y \in [0,1]^K$ and:
\begin{equation} \label{eq:aggregate_max}
l = - \sum_{k=1}^{K} y(k) \ln p_{\text{AT}}(k) + (1 - y(k)) \ln (1 - p_{\text{AT}}(k)).
\end{equation}

We then use the trained sound event detection model to perform framewise event prediction $ p_{\text{SED}} $ of an audio clip. We denote the anchor segment score of the $ k $-th sound class as:

\begin{align} \label{eq:sed_area}
q_{k}(t) = \sum_{t - \tau / 2}^{t + \tau / 2} p_{\text{SED}}(t, k),
\end{align}
where $ \tau $ is the duration of anchor segments. Then, the center time $t$ of the optimal anchor segment is obtained by:
\begin{align} \label{eq:argmax_sed_area}
t_{\text{anchor}} = \underset{t}{\text{argmax}} \ q_{k}(t).
\end{align}
The red block in Fig. \ref{fig:10s_audioset_speech} shows the detected anchor segment. We apply the anchor segment mining strategy as described in (\ref{eq:sed_area}) and (\ref{eq:argmax_sed_area}) process a mini-batch of the audio clips $ \{ x_{1}, ..., x_{B} \} $ into a mini-batch of anchor segments $ \{ s_{1}, ..., s_{B} \} $.

\begin{algorithm}[t]
	\caption{Prepare a mini-batch of data.}\label{alg:mining}
	\begin{algorithmic}[1]
            \State \textbf{Inputs:} dataset $ D =\{a_{n}, y_{n}\}_{n=1}^{N} $ containing $ K $ sound classes, mini batch size $ B $.
            \State \textbf{Outputs:} a mini-batch of mixture and anchor segment pairs $\{(x_{i}, s_{i})\}_{i=1}^{B}$.
            \State \textbf{Step 1: Balanced Sampling}: Uniformly sample $ B $ sound classes from sound classes $ \{1, ..., K\} $ without replacement.
            Sample one audio clip $a_{b}, b=1, ..., B$ for each selected sound class. We denote the mini-batch of audio clips as $ \{a_{1}, ..., a_{B}\} $.
            \State \textbf{Step 2: Anchor Segment Detection}:
            Apply a pretrained SED model on each $ a_{i} $ to detect the optimal anchor time stamp $t_{i}$ by (\ref{eq:sed_area})(\ref{eq:argmax_sed_area}). Extract the anchor segment $s_{i} \in \mathbb{R}^{L'}$ whose center is $t_{i}$.
            \State \textbf{Step 3 (optional)}: Use an audio tagging model to predict the event presence probability of $s_{i}$ and apply threshold $ \theta \in [0, 1]^{K} $ to get binary results $ r_{i} \in \{0, 1\}^{K} $ where $ r_{i}(k) $ is set to 1 if the presence probability is larger than $ \theta(k) $. Permute $ \{s_{i}\}_{i=1}^{B} $ so that $ r_{i} \wedge r_{(i+1)\%B} = \textbf{0} $, where $\%$ is the modulo operator.
            \State Create mixture source pairs $\{(x_{i}, s_{i})\}_{i=1}^{B}$ by $ x_{i} = s_{i} + s_{(i+1)\%B} $.
	\end{algorithmic}
\end{algorithm}

Algorithm \ref{alg:mining} shows the procedure for creating training data. Step 1 describes audio clip sampling and Step 2 describes anchor segment mining. To further avoid two anchor segments containing the same classes being mixed, we propose an optional Step 3 to mine anchor segments from a mini-batch of audio clips $ \{ s_{1}, ..., s_{B} \} $ to constitute mixtures. Step 4 describes mixing detected anchor segments into mixtures to train the USS system.


Fig. \ref{fig:pann-htsat} shows the model architectures of both PANNs and HTS-AT as two audio tagging models we employed. The DecisionLevel systems of PANNs \cite{kong2020panns} provide framewise predictions and contain VGG-like CNNs to convert an audio mel-spectrogram into feature maps. The model averages the feature maps over the time axis to obtain a final event classification vector. 
The framewise prediction $ p_{\text{SED}} \in [0, 1]^{T \times K}$ indicates the SED result. Additionally, the output of the penultimate layer with a size of $(T, H)$ can be used to obtain its averaged vector with a size of $H$ as a latent source embedding for the query-based source separation in our system, where $ H $ is the dimension of the latent embedding.

HTS-AT \cite{chen2022hts} is a hierarchical token-semantic transformer for audio classification. It applies Swin-Transformer \cite{liu2021swin} to an audio classification task. In the right of Fig. \ref{fig:pann-htsat}, a mel-spectrogram is cut into different patch tokens with a patch-embed CNN and sent into the Transformer in order. The time and frequency lengths of the patch are equal to $P \times P$, where $P$ is the patch size. To better capture the relationship between frequency bins of the same time frame, HTS-AT first splits the mel-spectrogram into windows $w_1, w_2, ..., w_n$ and then splits the patches in each window. The order of tokens $Q$ follows \textbf{time$\to$frequency$\to$window}. The patch tokens pass through several network groups, each of which contains several transformer-encoder blocks. Between every two groups, a patch-merge layer is applied to reduce the number of tokens to construct a hierarchical representation. Each transformer-encoder block is a Swin-transformer \cite{liu2021swin} block with the shifted window attention module, a modified self-attention module to improve the training efficiency. 

Then, HTS-AT applies a token-semantic 2D-CNN to further process the reshaped output $(\frac{T}{8P}, \frac{F}{8P}, H)$ into the framewise event presence map $(T,K)$ which can be averaged to an event classification vector $K$. The latent embedding, at the same time, is produced by averaging the reshaped output into a $H$-dimension vector with an average-pooling layer.

\subsection{Source Query Conditions} \label{section:source_embedding}

In contrast to previous query-based source separators that extract pre-defined representations \cite{kong2020source} or learnable representations \cite{chen2022zero, gfeller2021one, delcroix2022soundbeam} from clean sources, we propose to extract query embeddings from anchor segments to control sources to separate from a mixture. We introduce four types of embeddings: a hard one-hot embedding with a dimension of $ K $ where $K$ is the number of sound classes, a soft probability condition with a dimension of $ K $, a latent embedding condition with a dimension of $ H $, where $ H $ is the latent embedding dimension, and a learnable embedding condition with a dimension of $ H $.

\subsubsection{Hard One-Hot Condition}
We define the hard one-hot condition of an anchor segment $ s_{i} $ as $ c_{i} \in [0,1]^K$, where $ c_{i} $ is the one-hot representation of tags of the audio clip $ x_{i} $. The hard one-hot condition has been used in music source separation \cite{meseguer2019conditioned}. Hard one-hot embedding requires clean source data for training source separation systems.

\subsubsection{Soft Probability Condition}
The soft probability condition applies pretrained audio tagging models, such as PANNs or HTS-AT to calculate the event classification probability $ c_{i} = p_{\text{AT}}(s_{i}) $ of an anchor segment as the query embedding. For the weakly labelled dataset, the soft probability condition provides a continuous value prediction of what sounds are in an anchor segment than the hard one-hot condition. The advantage of the soft probability condition is that it explicitly presents the SED result. 




\subsubsection{Latent Embedding Condition}
The latent embedding with a dimension of $ H $ is calculated from the penultimate layer of an audio tagging model. The advantage of using the latent embedding is that the separation is not limited to the given $ K $ sound classes. The USS system can be used to separate arbitrary sound classes with a query embedding as input, allowing us to achieve USS. We investigate a variety of PANNs, including CNN5, CNN10, CNN14, and an HTS-AT model to extract the latent embeddings and we evaluate their efficiency on the separation performance. 
We denote the embedding condition extraction of as $c_{i} = f_{\text{emb}}(s_{i})$.

\subsubsection{Learnable Condition}
The latent embedding condition can be learned during the training of our universal source separation system. The first method is to fine-tune the parameters of the query net. The second method is to freeze the parameters of the query net and add a cloned query net containing learnable parameters as a shortcut branch $ f_{\text{shortcut}} $ to construct the embedding. The third method is to add learnable fully connected layers $f_{\text{ada}}(\cdot)$ on top of the query net where ada is the abbreviation for adaptive. The embedding can be extracted by: $c_{i}=f_{\text{ada}}(f_{\text{AT}}(s_{i}), \theta)$. 

\subsection{Query-based Source Separation} \label{section:query_ssp}
A typical source separator is a single-input-single-output model \cite{luo2021rethinking} that deals with one specific source, such as vocal, drum, or bass. To enable the model to separate arbitrary sound sources, we apply a query-based source separator by introducing the conditional embeddings as described in Section \ref{section:source_embedding} into the ResUNet source separation backbone \cite{kong2021decoupling} to build a single-input single-output source separation model \cite{luo2021rethinking}.

As mentioned in Section \ref{section:ss_strong}, the input to the ResUNet separator is a mixture of audio segments. First, we apply a short-time Fourier transform (STFT) to the waveform to extract the complex spectrum $ X = \mathbb{C}^{T \times F} $. Then, we follow the same setting of \cite{kong2021decoupling} to construct an encoder-decoder network to process the magnitude spectrogram $ |X| $.  The ResUNet encoder-decoder consists of 6 encoder blocks, 4 bottleneck blocks, and 6 decoder blocks. Each encoder block consists of 4 residual convolutional blocks to downsample the spectrogram into a bottleneck feature, and each decoder block consists of 4 residual deconvolutional blocks to upsample the feature back to separation components. The skip-connection is applied from each encoder block to the corresponding decoder block of the same downsampling/upsampling rate. The residual block contains 2 convolutional layers, 2 batch normalization \cite{ioffe2015batch} layers, and 2 Leaky-ReLU activation layers. An additional residual shortcut is added between the input and the output of each residual block. The details of the model architecture can be found at \cite{kong2021decoupling}.

The ResUNet separator outputs the magnitudes and the phases of the cIRM $ M \in \mathbb{C}^{T \times F} $. The separated complex spectrum can be obtained by:
\begin{equation} \label{eq:mask_mul}
\begin{split}
\hat{S} &= M \odot X \\
& = |M| \odot |X|e^{j (\angle M + \angle X)},
\end{split}
\end{equation}
where both $|M|$ and $\angle M$ are calculated from the output of the separator. The separated source can be obtained by multiplying the STFT of the mixture by the cIRM $ M $. The complex multiplication can also be decoupled into a magnitude multiplication part and a phase addition part. The magnitude $ |M| $ controls how much the magnitude of $ |X| $ should be scaled, and the angle $ \angle M $ controls how much the angle of $ X $ should be rotated. 

Based on the ResUNet separator, we adopt a feature-wise linear modulation (FiLM) \cite{perez2018film} method to construct convolutional blocks within the separator. 
We apply a pre-activation architecture \cite{he2016identity} for all of the encoder and decoder layers, we incorporate the conditional embedding as: 
\begin{equation} \label{eq:FiLM}
h^{l} = W * (\sigma(\text{BN}(h^{l-1}) + Vc))
\end{equation}
where $h^{l-1}$ is the feature map of the $l-1$-th layer, and $ V $ is a fully connected layer to map the conditional vector $ c $ into an embedding space. $Vc$ modulates the value $BN(h^{l-1})$. The value is convolved with the weight $W$ to output $h^{l}$. 
The training of our weakly labeled USS system is illustrated in Algorithm \ref{alg:uss}. 



	        

\begin{algorithm}[t]
	\caption{Training of a USS system.}\label{alg:uss}
	\begin{algorithmic}[1]
	    \State \textbf{Inputs:} Dataset $D$, e.g., AudioSet.
            \State \textbf{Outputs:} A trained USS model.
	    \While {loss function $l$ does not converge}
            \State prepare a mini-batch of mixture source training pairs $ \{(x_{i}, s_{i})\}_{i=1}^{B} $ by Algorithm \ref{alg:mining}
	    \State Calculate source query embeddings $ \{c_{i}\}_{i=1}^{B} $ by any of query nets as described in Section \ref{section:source_embedding}.
            \For {\text{each} $s_i$}
                \State Obtain the separation $\hat{s_{i}}=f(x_{i},c_i)$
                \State Calculate loss by $ l(\hat{s_{i}}, s_{i}) $
            \EndFor
	    \EndWhile
	\end{algorithmic}
\end{algorithm}

\subsection{Data augmentation}\label{section:data_augmentation}
When constituting the mixture with $ s_{i} $ and $ s_{i+1} $, the amplitude of $ s_{i} $ and $ s_{i+1} $ can be different. We propose an energy augmentation to augment data. That is, we first calculate the energy of a signal $ s_{i} $ by $ E = || s_{i} ||_{2}^{2} $. We denote the energy of $ s_{i} $ and $ s_{i+1} $ as $ E_{i} $ and $ E_{i+1} $. We apply a scaling factor $ \alpha_{i} = \sqrt{E_{i} / E_{i+1}} $ to $ s_{i+1} $ when creating the mixture $ x_{i} $:

\begin{equation} \label{eq:scale_augmentation}
x = s_{i} + \alpha s_{i+1}.
\end{equation}

\noindent By this means, both anchor segments $ s_{i} $ and $ s_{i+1} $ have the same energy which is beneficial to the optimization of the separation system. We will show this in Section \ref{section:exp_data_augmentation}. On the one hand, we match the energy of anchor segments to let the neural network learn to separate the sound classes. On the other hand, the amplitude diversity of sound classes is increased. We will show this energy augmentation is beneficial in our experiments. 

\subsection{Loss functions}
We propose to use the L1 loss between the predicted and ground truth waveforms following \cite{kong2021decoupling} to train the end-to-end universal source separation system:

\begin{equation} \label{eq:loss}
l = ||s - \hat{s}||_{1},
\end{equation}

\noindent where $ l $ is the loss function used to train the neural network. A lower loss in (\ref{eq:loss}) indicates that the separated signal $ \hat{s} $ is closer to the ground truth signal $ s $. In training, the gradients of parameters are calculated by $ \partial l / \partial \theta $, where $ \theta $ are the parameters of the neural network. 

\subsection{Inference}
In training, the oracle embedding of an anchor segment can be calculated by $ f_{\text{emb}}(s_{i}) $. In inference, for the hard one-hot condition and soft probability condition, we can simply use 1) the one-hot representation of the $ k $-th sound class to separate the audio of the $k$-th sound class. 2) Only remaining soft probabilities' $ \{k_{j}\}_{j=1}^{J} $ indexes values as the condition, where $ J $ is the number of sound classes to separate. 

However, for the latent embedding condition and learnable condition, we need to calculate the embedding $ c $ from the training dataset by: 
\begin{equation} \label{eq:avg_emb}
c = \frac{1}{N}\sum_{n=1}^{N}f_{\text{emb}}(s_{n})
\end{equation}
\noindent where $ \{s_{n}\}_{n=1}^{N} $ are query samples of one sound class and $ N $ is the number of query samples. That is, we average all conditional embeddings of query samples from the same sound class to constitute $ c $.






\begin{algorithm}[t]
	\caption{Automatic sound event detection hierarchical USS.}\label{alg:automatic_filt_sep}
	\begin{algorithmic}[1]
            \State \textbf{Inputs:} An arbitrary duration audio clip $ x $. A trained USS system $ f_{\text{SS}} $. A trained audio tagging system $ f_{\text{AT}} $. Hierarchical level $l$.
            \State \textbf{Outputs:} Separated sources $ O = \{\hat{s}_{j}\}_{j \in C} $ where $ C $ is the indexes .
            \State Split $ x $ into non-overlapped short segments $\{x_{i}\}_{i=1}^{I}$ where $I$ is the number of segments. 
            \State Apply $ f_{\text{AT}} $ on all segments to obtain $ P(i, k) $ with a size of $ I \times K$, where $K$ is the number of sound classes.
            \State\textit{\# Calculate ontology predictions.}
            \If{hierarchical\_separation}
            \State $ Q(i, j) = \text{HierarchicalOntologyGrouping}(P(i, k), l) $ following Algorithm \ref{alg:ontology_grouping}. $ Q(i, j) $ has a shape of $ I \times J $ where $ J $ is the number of sound classes in the $l$-th level.
            \EndIf
            \State\textit{\# Detect active sound event.}
            \State $ C = \{\} $  \text{\# \textit{Active sound class indexes.}}
            \For {$j=1, ..., J$}
                \If{$ \text{max}_{i}Q(j, j) > \delta$}: 
                    \State $ C = C \cup \{j\} $
                \EndIf
            \EndFor
            \State\textit{\# Do separation.}
            \State $ O = \{\} $ \textit{\# Separated sources of active sound classes.}
            \For {$ j \in C $}
                \For {$i=1, ..., I$}
                    \If{$ Q(i, j) > \delta$}:
                        \State Get condition $ c_{j} $ by (\ref{eq:inference_hierarchy}).
                        \State $ \hat{s}_{ij} = f_{\text{ss}}(x_{i}, c_{j}) $
                    \Else
                        \State $ \hat{s}_{ij} = \textbf{0} $
                    \EndIf
                \EndFor
                \State $\hat{s}_{j} = \{\hat{s}_{ij}\}_{i=1}^{I}$
                \State $ O = O \cup \{\hat{s}_{j}\} $
            \EndFor
	\end{algorithmic}
\end{algorithm}

\begin{algorithm}[t]
	\caption{Hierarchical Ontology Grouping.}\label{alg:ontology_grouping}
	\begin{algorithmic}[1]
        \State \textbf{Inputs:} segment-wise prediction $ P(i, k) $ with a size of $ I \times K $, hierarchy level $ l $.
        \State \textbf{Outputs:} $ Q(i, j) $ with a size of $ I \times J $ where $ J $ is the number of children sound classes of the $l$-th ontology level.
        \For{$ j \in 1, ..., J $}
            \State $ Q(i, j) = \text{max}_{k}\{P(i, k)\}_{k \in \text{children}(j)} $
        \EndFor
	\end{algorithmic}
\end{algorithm}

\subsection{Inference with Hierarchical AudioSet Ontology}
\label{section:hierarchy_separation}
We propose a hierarchical separation strategy to address the USS problem. It is usually unknown how many and what sound classes are present and which need to be separated in an audio clip. To address this problem, we propose a hierarchical sound class detection strategy to detect the sound classes presence. We separate those sound classes by using the trained USS system. 

Algorithm \ref{alg:automatic_filt_sep} shows the automatic sound class detection and separation steps. The input to the USS system is an audio clip $ x $. We first split the audio clip into short segments and apply an audio tagging system $ f_{\text{AT}} $ to calculate the segment-wise prediction $ P(t, k) $ with a size of $ I \times K $, where $ I $ and $K$ are the number of segments and sound classes, respectively. The AudioSet ontology has a tree structure. The first level of the tree structure contains seven sound classes described in the AudioSet ontology, including ``Human sounds'', ``Animal'', ``Music'', ``Source-ambiguous sounds'', ``Sounds of things'', ``Nature sounds'', and ``Channel, environment and background''. Each root category contains several sub-level sound classes. The second level and the third levels contain 41 and 251 sound classes, respectively, as described in the AudioSet ontology \cite{gemmeke2017audio}. The tree structure has a maximum depth of six levels. 

In inference, the USS system supports hierarchical source separation with different levels. We denote the sound classes of level $ l $ as $ C = \{c_{j}\}_{j=1}^{J} $, where $ J $ is the number of sound classes in the $l$-th level. For a sound class $ j $ in the $l$-th level, we denote the set of all its children's sound classes as $ \text{children}(j) $ including $j$. For example, for the human sounds class $ 0 $, there are $ \text{children}(0) = \{0, 1, ..., 72\} $. 
We set score $ Q(i, j) = \text{max}_{k}\{P(i, j)\}_{k\in \text{children}(j)} $. We detect a sound class $j$ as active if $ \text{max}_{i}Q(i, j) $ larger than a threshold $ \theta $. We set separated segments to silence if $ Q(i, j) $ is smaller than $ \theta $. Then, we apply the USS by using (\ref{eq:inference_hierarchy}) as the condition:
\begin{equation} \label{eq:inference_hierarchy}
c_{k}=\left\{\begin{matrix}
f_{\text{AT}}(x), k \in \text{children}(j)\\ 
0,k \notin \text{children}(j).
\end{matrix}\right.
\end{equation}
The USS procedure is described in Algorithm \ref{alg:ontology_grouping}. 


\section{Experiments}
In this section, we investigate our proposed universal source separation system on several tasks, including AudioSet separation \cite{gemmeke2017audio}, sound event separation \cite{fonseca2018general, fonseca2020fsd50k}, music source separation \cite{rafii2017musdb18, manilow2019cutting}, and speech enhancement \cite{veaux2013voice}. Our USS system is trained only on the large-scale weakly labelled AudioSet \cite{gemmeke2017audio} without using any clean training data, which is a major difference from the previous source separation systems that are trained on specific datasets with clean sources \cite{jansson2017singing, chandna2017monoaural, stoter2019open, takahashi2018mmdenselstm, hennequin2020spleeter}. The trained USS system can address a wide range of source separation tasks without being finetuned. 

\subsection{Training Dataset}
AudioSet is a large-scale weakly labelled audio dataset containing 2 million 10-second audio clips sourced from the YouTube website. Audio clips are only labelled with the presence or absence of sound classes, without knowing when the sound events occur. There are 527 sound classes in its released version, covering a wide range of sound classes in the world, such as ``Human sounds'', ``Animal'', etc. The training set consists of 2,063,839 audio clips, including a balanced subset of 22,160 audio clips. There are at least 50 audio clips for each sound class in the balanced training set. 
Although some audio links are no longer available, we successfully downloaded 1,934,187 (94\%) audio clips from the full training set.
All audio clips are padded with silence into 10 seconds. Due to the fact that a large amount of audio recordings from YouTube have sampling rates lower than 32 kHz, we resample all audio recordings into mono and 32 kHz. 

\subsection{Training Details}
We select anchor segments as described in Section \ref{section:anchor} and mix two anchor segments to constitute a mixture $ x $. The duration of each anchor segment is 2 seconds. We investigate different anchor segment durations in Section \ref{section:Anchor_segment_duration}. We apply matching energy data augmentation as described in Section \ref{section:data_augmentation} to scale two anchor segments to have the same energy, and extract the short-time Fourier transform (STFT) feature $ X $ from $ x $ with a Hann window size of 1024 and a hop size 320. This hop size leads to 100 frames in a second consistent to the audio tagging systems in PANNs \cite{kong2020panns} and HTS-AT \cite{chen2022hts}.

The query net is a CNN14 of PANNs or HTS-AT. The query net is pretrained on the AudioSet tagging task \cite{gemmeke2017audio, chen2022hts} and the parameters are frozen during the training of the USS system. The prediction and the embedding layers of the query net have dimensions of 527 and 2048, respectively. Either the prediction layer or the embedding layer is connected to fully connected layers and input to all layers of the source separation branch as FiLMs. We adopt ResUNet \cite{kong2021decoupling} as the source separation branch. The 30-layer ResUNet consists of 6 encoder and 6 decoder blocks. Each encoder block consists of two convolutional layers with kernel sizes of $ 3 \times 3 $. Following the pre-activation strategy \cite{he2016identity}, we apply batch normalization \cite{ioffe2015batch} and leaky ReLU \cite{xu2015empirical} before each convolutional layer. The FiLM is added to each convolutional layer as described in (\ref{eq:FiLM}). The number of output feature maps of the encoder blocks are 32, 64, 128, 256, 512, and 1024, respectively. The decoder blocks are symmetric to the encoder blocks. We apply an Adam optimizer \cite{kingma2014adam} with a learning rate of $ 10^{-3} $ to train the system. A batch size of 16 is used to train the USS system. The total training steps is 600 k trained for 3 days on a single Tesla V100 GPU card.

\subsection{Conditional Embedding Calculation}
For AudioSet source separation, the oracle embedding or each anchor segment is calculated by:
\begin{equation} \label{eq:ora_emb}
c = f_{\text{emb}}(s)
\end{equation}
where $ s $ is the clean source. Using oracle embedding as condition indicates the upper bound of the universal source separation system. For real applications, we calculate the conditional embeddings by (\ref{eq:avg_emb}) from the training set of the AudioSet, FSD50Kaggle2018, FSD50k, MUSDB18, Slakkh2100, and VoicebankDemand datasets to evaluate on those datasets, respectively.

\begin{table*}
  \caption{USS results with different conditional embedding types.}
  \label{tab:emb_types}
  \centering
  \resizebox{\textwidth}{!}{
  \begin{tabular}{lccccccccccccccccccc}
    \toprule
    & \multicolumn{2}{c}{\textbf{AudioSet (SDRi)}} & \multicolumn{2}{c}{\textbf{FSDK2018}} & \multicolumn{2}{c}{\textbf{FSD50k}} & \multicolumn{4}{c}{\textbf{MUSDB18}} & \multicolumn{2}{c}{\textbf{Slakh2100}} & \multicolumn{2}{c}{\textbf{VoicebankDemand}} \\
	\cmidrule(lr){2-3} \cmidrule(lr){4-5} \cmidrule(lr){6-7}\cmidrule(lr){8-11}\cmidrule(lr){12-13}\cmidrule(lr){14-15}
    & ora. emb & avg. emb & SDR & SDRi & SDR & SDRi & SDR & SDRi & wSDR & wSDRi & SDR & SDRi & PESQ & SSNR  \\
    \midrule
    wav2vec (46c) & 8.87 & 4.30 & 8.95 & 8.91 & 4.52 & 4.70 & 1.90 & 7.03 & 2.96 & 8.37 & -1.08 & 6.66 & 2.11 & 6.02 \\
    speaker (46d) & 8.87 & 2.82 & 6.69 & 6.65 & 3.00 & 3.03 & 1.52 & 6.85 & 2.48 & 7.94 & 0.18 & 7.92 & 2.13 & 4.72 \\
    \midrule
    CNN6 (45a2) & 8.68 & 5.30 & 10.36 & 10.31 & 5.25 & 5.50 & 3.05 & 8.43 & 3.94 & 9.42 & -0.37 & 7.37 & 2.27 & 9.39 \\
    CNN10 (45a3) & 8.35 & 5.36 & 9.95 & 9.90 & 5.19 & 5.43 & 2.87 & 8.10 & 4.11 & 9.34 & -0.27 & 7.47 & 2.27 & 8.68 \\
    +CNN14 (44a) & 8.26 & 5.57 & 10.61 & 10.57 & 5.54 & 5.79 & 3.08 & 8.12 & 4.02 & 9.22 & -0.46 & 7.28 & 2.18 & 9.00 \\
    HTSAT (45c) & 9.38 & 3.78 & 7.95 & 7.91 & 3.38 & 3.51 & 2.83 & 8.48 & 3.77 & 9.36 & 0.81 & 8.55 & 2.23 & 8.78 \\
    \bottomrule
\end{tabular}
}
\end{table*}

\begin{table*}
  \caption{USS results with soft audio tagging and latent embedding as condition.}
  \label{table:seg_pred_emb}
  \centering
  \resizebox{\textwidth}{!}{%
  \begin{tabular}{lccccccccccccccccccc}
    \toprule
    & \multicolumn{2}{c}{\textbf{AudioSet (SDRi)}} & \multicolumn{2}{c}{\textbf{FSDK2018}} & \multicolumn{2}{c}{\textbf{FSD50k}} & \multicolumn{4}{c}{\textbf{MUSDB18}} & \multicolumn{2}{c}{\textbf{Slakh2100}} & \multicolumn{2}{c}{\textbf{VoicebankDemand}} \\
	\cmidrule(lr){2-3} \cmidrule(lr){4-5} \cmidrule(lr){6-7}\cmidrule(lr){8-11}\cmidrule(lr){12-13}\cmidrule(lr){14-15}
    & ora. emb & avg. emb & SDR & SDRi & SDR & SDRi & SDR & SDRi & wSDR & wSDRi & SDR & SDRi & PESQ & SSNR  \\
    \midrule
    Segment prediction (dim=527) (46b) & 7.80 & 6.42 & 11.22 & 11.18 & 6.60 & 6.92 & 2.48 & 7.26 & 3.58 & 8.80 & -1.69 & 6.05 & 2.20 & 8.30 \\
    +Embedding (dim=2048) & 8.26 & 5.57 & 10.61 & 10.57 & 5.54 & 5.79 & 3.08 & 8.12 & 4.02 & 9.22 & -0.46 & 7.28 & 2.18 & 9.00 \\
    \bottomrule
\end{tabular}}
\end{table*}

\subsection{Evaluation Datasets}
\subsubsection{AudioSet} The evaluation set of AudioSet \cite{gemmeke2017audio} contains 20,317 audio clips with 527 sound classes. We successfully downloaded 18,887 out of 20,317 (93\%) audio clips from the evaluation set. AudioSet source separation is a challenging problem due to USS need to separate 527 sound classes using a single model. We are the first to propose using AudioSet \cite{kong2020source} to evaluate the USS. To create evaluation data, similarly to Section \ref{section:anchor}, we first apply a sound event detection system to each 10-second audio clip to detect anchor segments. Then, we select two anchor segments from different sound classes and sum them as a mixture for evaluation. We create 100 mixtures for each sound class, leading to 52,700 mixtures for all sound classes in total.

\subsubsection{FSDKaggle2018}
The FSDKaggle2018 \cite{fonseca2018general} is a general-purpose audio tagging dataset containing 41 sound classes ranging from musical instruments, human sounds, domestic sounds, and animals, etc. The duration of the audio clips ranges from 300 ms to 30 s. Each audio clip contains a unique audio tag. The test set is composed of 1,600 audio clips with manually-verified annotations. We pad or truncate each audio clip into 2-second segment from the start, considering sound events usually occur in the start of audio clips. We mix two segments from different sound classes to consist a pair. We constitute 100 mixtures for each sound class. This leads to a total of 4,100 evaluation pairs. 

\subsubsection{FSD50K dataset}
The Freesound Dataset 50k (FSD50K) dataset \cite{fonseca2020fsd50k} contains 51,197 training clips distributed in 200 sound classes from the AudioSet ontology. In contrast to the FSDKaggle2018 dataset, each audio clip may contain multiple tags with a hierarchical architecture. There are an average of 2.79 tags in each audio clip. All audio clips are sourced from Freesound\footnote{\url{https://freesound.org/}}. There are 10,231 audio clips distributed in 195 sound classes in the test set. Audio clips have variable durations between 0.3s to 30s, with an average duration of 7.1 seconds. We mix two segments from different sound classes to consist a pair. We create 100 mixtures for each sound class. This leads to in total 19,500 evaluation pairs. 

\subsubsection{MUSDB18}
The MUSDB18 dataset \cite{rafii2017musdb18} is designed for the music source separation task. The test set of the MUSDB18 dataset contains 50 songs with four types of stems, including vocals, bass, drums, and other. We linearly sum all stems to constitute mixtures as input to the USS system. We use the museval toolkit\footnote{https://github.com/sigsep/sigsep-mus-eval} to evaluate the SDR metrics. 

\subsubsection{Slakh2100}
The Slakh2100 dataset \cite{manilow2019cutting} is a multiple-instrument dataset for music source separation and transcription. The test of the Slakh2100 dataset contains 225 songs. The sound of different instruments are rendered by 167 different types of plugins. We filtered 151 non-silent plugin types for evaluation. Different from the MUSDB18 dataset, there can be over 10 instruments in a song, leading to the Slakh2100 instrument separation a challenging problem. 

\subsubsection{Voicebank-Demand}
The Voicebank-Demand \cite{veaux2013voice} dataset is designed for the speech enhancement task. The Voicebank dataset \cite{veaux2013voice} contains clean speech. The Demand dataset \cite{thiemann2013demand} contains multiple different background sounds that are used to create mixtures. The noisy utterances are created by mixing the VoiceBank dataset and the Demand dataset under signal-to-noise ratios of 15, 10, 5, and 0 dB. The test set of the Voicebank-Demand dataset contains 824 utterances in total.

\subsection{Evaluation Metrics}
We use the signal-to-distortion ratio (SDR) \cite{vincent2006performance} and SDR improvement (SDRi) \cite{vincent2006performance} to evaluate the source separation performance. The SDR is defined as:
\begin{equation} \label{eq:sdr}
\text{SDR}(s, \hat{s}) = 10 \text{log}_{10} \left ( \frac{||s||^2}{||s - \hat{s}||^2} \right )
\end{equation}
\noindent where $ s $ and $ \hat{s} $ are the target source and estimated source, respectively. Larger SDR indicates better separation performance. The SDRi is proposed to evaluate how much SDR a USS system improves compared to without separation:
\begin{equation} \label{eq:sdr}
\text{SDRi} = \text{SDR}(s, \hat{s}) - \text{SDR}(s, x)
\end{equation}
\noindent where $ x $ is the mixture signal. For the speech enhancement task, we apply the Perceptual evaluation of speech quality (PESQ) \cite{recommendation2001perceptual} and segmental signal-to-ratio noise (SSNR) \cite{quackenbush1988objective} for evaluation. 

\begin{table*}
  \caption{USS results with freeze, finetune, and adapt conditional embeddings.}
  \label{tab:freeze_fintune_emb}
  \centering
  \resizebox{\textwidth}{!}{%
  \begin{tabular}{lccccccccccccccccccc}
    \toprule
    & \multicolumn{2}{c}{\textbf{AudioSet (SDRi)}} & \multicolumn{2}{c}{\textbf{FSDK2018}} & \multicolumn{2}{c}{\textbf{FSD50k}} & \multicolumn{4}{c}{\textbf{MUSDB18}} & \multicolumn{2}{c}{\textbf{Slakh2100}} & \multicolumn{2}{c}{\textbf{VoicebankDemand}} \\
	\cmidrule(lr){2-3} \cmidrule(lr){4-5} \cmidrule(lr){6-7}\cmidrule(lr){8-11}\cmidrule(lr){12-13}\cmidrule(lr){14-15}
    & ora. emb & avg. emb & SDR & SDRi & SDR & SDRi & SDR & SDRi & wSDR & wSDRi & SDR & SDRi & PESQ & SSNR  \\
    \midrule
    CNN14 (random weights) & 8.51 & 2.82 & 5.96 & 5.91 & 2.82 & 2.82 & -0.48 & 4.59 & 1.97 & 7.08 & -1.34 & 6.40 & 2.28 & 6.96 \\
    CNN14 (scratch) & 2.38 &  2.38 & 2.46 & 2.41 & 2.22 & 2.15 & 0.71 & 5.78 & 1.16 & 6.30 & -1.20 & 6.54 & 1.62 & -0.28 \\
    CNN14 (finetune) & 9.83 & 1.96 & 3.42 & 3.38 & 1.50 & 1.40 & 2.10 & 7.77 & 3.10 & 8.52 & 1.39 & 9.12 & 1.77 & 3.52 \\
    +CNN14 (freeze) & 8.26 & 5.57 & 10.61 & 10.57 & 5.54 & 5.79 & 3.08 & 8.12 & 4.02 & 9.22 & -0.46 & 7.28 & 2.18 & 9.00 \\
    CNN14 + shortcut & 6.95 & 4.57 & 9.29 & 9.25 & 4.74 & 4.94 & 1.84 & 7.05 & 3.40 & 8.78 & -1.44 & 6.30 & 2.06 & 8.91 \\
    CNN14 + adaptor & 8.01 & 5.81 & 11.00 & 10.96 & 5.79 & 6.07 & 2.95 & 7.96 & 3.90 & 9.24 & -0.87 & 6.87 & 2.30 & 9.60 \\
    \bottomrule
\end{tabular}}
\end{table*}

\begin{table*}
  \caption{USS results with different backbone models.}
  \label{table:models}
  \centering
  \resizebox{\textwidth}{!}{
  \begin{tabular}{lccccccccccccccccccc}
    \toprule
    & \multicolumn{2}{c}{\textbf{AudioSet (SDRi)}} & \multicolumn{2}{c}{\textbf{FSDK2018}} & \multicolumn{2}{c}{\textbf{FSD50k}} & \multicolumn{4}{c}{\textbf{MUSDB18}} & \multicolumn{2}{c}{\textbf{Slakh2100}} & \multicolumn{2}{c}{\textbf{VoicebankDemand}} \\
	\cmidrule(lr){2-3} \cmidrule(lr){4-5} \cmidrule(lr){6-7}\cmidrule(lr){8-11}\cmidrule(lr){12-13}\cmidrule(lr){14-15}
    & ora. emb & avg. emb & SDR & SDRi & SDR & SDRi & SDR & SDRi & wSDR & wSDRi & SDR & SDRi & PESQ & SSNR  \\
    \midrule
    open-unmix & 3.96 & 3.39 & 3.90 & 3.86 & 2.96 & 2.92 & 0.40 & 5.50 & 2.13 & 7.42 & -1.09 & 6.65 & 2.40 & 2.26 \\
    ConvTasNet & 6.96 & 5.00 & 9.49 & 9.45 & 5.31 & 5.54 & 0.61 & 5.61 & 2.64 & 8.10 & -2.96 & 4.77 & 1.87 & 6.46 \\
    UNet & 8.14 & 5.50 & 10.83 & 10.79 & 5.49 & 5.75 & 2.49 & 7.78 & 3.70 & 9.17 & -0.45 & 7.29 & 2.12 & 8.47 \\ 
    ResUNet30 & 8.26 & 5.57 & 10.61 & 10.57 & 5.54 & 5.79 & 3.08 & 8.12 & 4.02 & 9.22 & -0.46 & 7.28 & 2.18 & 9.00 \\
    ResUNet60 & 7.97 & 5.70 & 11.34 & 11.30 & 6.04 & 6.32 & 1.71 & 6.81 & 3.64 & 9.01 & -2.77 & 4.97 & 2.40 & 9.35 \\
    \bottomrule
\end{tabular}
}
\end{table*}

\subsection{Results Analysis}
\subsubsection{Conditional Embedding Types}
The default configuration of our USS system is a 30-layer ResUNet30 trained on the balanced set of AudioSet. Table \ref{tab:emb_types} shows the USS system results trained with different conditional embedding types including wav2vec \cite{schneider2019wav2vec}, speaker embeddings\footnote{https://github.com/RF5/simple-speaker-embedding}, CNN6, CNN10, CNN14 from PANNs \cite{kong2020panns}, and HTS-AT \cite{chen2022hts}. The wav2vec embedding is trained using unsupervised contrastive learning on 960 hours of speech data. The wav2vec embedding is averaged along the time axis to a single embedding with a dimension of 512. The speaker embedding is a gated recurrent unit (GRU) with three recurrent layers operates on log mel-spectrogram and has output has a shape of 256. The CNN6 and the CNN10 have dimensions of 512. The CNN14 and the HTS-AT have dimensions of 2048. The oracle embedding (ora emb) shows the results using (\ref{eq:ora_emb}) as condition. The average embedding (avg emb) shows the results using (\ref{eq:avg_emb}) as condition.

Table \ref{tab:emb_types} shows that the CNN6, CNN10, CNN14 embeddings achieve AudioSet SDR between 5.30 dB and 5.57 dB using the average embedding, outperforming the wav2vec of 4.30 dB and the speaker embedding of 2.82 dB. One possible explanation is that both wav2vec and the speaker embeddings are trained on speech data only, so that they are not comparable to PANNs and HTS-AT trained for general audio tagging. The wav2vec embedding slightly outperforms the speaker embedding on FSDKaggle2018, FSD50k, and MUSDB18 separation, indicating that the unsupervised learned ASR embeddings are more suitable for universal source separation. The HTS-AT achieves the highest oracle embedding SDR among all systems.  All of CNN6, CNN10, CNN14, and HTS-AT outperform the wav2vec embedding and the speaker embedding in AudioSet, FSDKaggle2018, FSD50k, MUSDB18, Slakh2100, and Voicebank-Demand datasets by a large margin. The CNN14 slightly outperforms CNN6 and CNN10. In the following experiments, we use CNN14 as the default conditional embedding.

\begin{table*}
  \caption{USS results trained with different anchor segment durations.}
  \label{table:segment_seconds}
  \centering
  \resizebox{\textwidth}{!}{
  \begin{tabular}{lccccccccccccccccccc}
    \toprule
    & \multicolumn{2}{c}{\textbf{AudioSet (SDRi)}} & \multicolumn{2}{c}{\textbf{FSDK2018}} & \multicolumn{2}{c}{\textbf{FSD50k}} & \multicolumn{4}{c}{\textbf{MUSDB18}} & \multicolumn{2}{c}{\textbf{Slakh2100}} & \multicolumn{2}{c}{\textbf{VoicebankDemand}} \\
	\cmidrule(lr){2-3} \cmidrule(lr){4-5} \cmidrule(lr){6-7}\cmidrule(lr){8-11}\cmidrule(lr){12-13}\cmidrule(lr){14-15}
    & ora. emb & avg. emb & SDR & SDRi & SDR & SDRi & SDR & SDRi & wSDR & wSDRi & SDR & SDRi & PESQ & SSNR  \\
    \midrule
    0.5 s & 4.07 & 2.86 & 4.51 & 4.47 & 2.55 & 2.51 & 0.97 & 0.78 & 2.61 & 2.43 & -0.79 & 6.95 & 1.57 & 5.96\\
    1s & 7.50 & 4.99 & 9.45 & 9.41 & 4.81 & 5.00 & 0.18 & -0.02 & 2.54 & 2.50 & -1.66 & 6.08 & 2.17 & 8.55\\
    +2s & 8.26 & 5.57 & 10.61 & 10.57 & 5.54 & 5.79 & 3.08 & 8.12 & 4.02 & 9.22 & -0.46 & 7.28 & 2.18 & 9.00 \\
    4s & 7.39 & 5.21 & 10.22 & 10.17 & 5.38 & 5.60 & 1.83 & 6.79 & 3.38 & 8.68 & -2.62 & 5.12 & -2.62 & 5.12 \\
    6s & 6.39 & 4.68 & 9.20 & 9.16 & 5.05 & 5.24 & 0.00 & 4.98 & 2.70 & 7.97 & -4.26 & 3.48 & 2.21 & 2.56\\
    8s & 6.26 & 4.48 & 8.85 & 8.80 & 4.77 & 4.94 & -3.67 & -4.00 & 1.60 & 1.50 & -5.68 & 2.06 & 2.24 & 2.35 \\
    10s & 6.29 & 4.47 & 9.11 & 9.07 & 4.80 & 4.98 & -2.68 & -2.79 & 1.56 & 1.53 & -5.07 & 2.67 & 2.13 & 2.14 \\
    \bottomrule
\end{tabular}
}
\end{table*}

\begin{table*}
  \caption{USS results with different anchor mining strategies.}
  \label{table:anchor_mining_strategy}
  \centering
  \resizebox{\textwidth}{!}{
  \begin{tabular}{lccccccccccccccccccc}
    \toprule
    & \multicolumn{2}{c}{\textbf{AudioSet (SDRi)}} & \multicolumn{2}{c}{\textbf{FSDK2018}} & \multicolumn{2}{c}{\textbf{FSD50k}} & \multicolumn{4}{c}{\textbf{MUSDB18}} & \multicolumn{2}{c}{\textbf{Slakh2100}} & \multicolumn{2}{c}{\textbf{VoicebankDemand}} \\
	\cmidrule(lr){2-3} \cmidrule(lr){4-5} \cmidrule(lr){6-7}\cmidrule(lr){8-11}\cmidrule(lr){12-13}\cmidrule(lr){14-15}
    & ora. emb & avg. emb & SDR & SDRi & SDR & SDRi & SDR & SDRi & wSDR & wSDRi & SDR & SDRi & PESQ & SSNR  \\
    \midrule
    mining & 8.26 & 5.57 & 10.61 & 10.57 & 5.54 & 5.79 & 3.08 & 8.12 & 4.02 & 9.22 & -0.46 & 7.28 & 2.18 & 9.00 \\
    in-clip random & 4.89 & 3.94 & 5.53 & 5.49 & 3.63 & 3.66 & 1.10 & 6.05 & 2.36 & 7.79 & -1.72 & 6.01 & 2.21 & 5.69 \\
    out-clip random & 8.19 & 5.90 & 11.06 & 11.01 & 6.04 & 6.34 & 2.57 & 7.68 & 3.81 & 9.17 & -1.08 & 6.66 & 2.39 & 9.48 \\
    \bottomrule
\end{tabular}
}
\end{table*}

Table \ref{table:seg_pred_emb} shows the comparison between using the CNN14 segment prediction with a dimension of 527 and the CNN14 embedding condition with a dimension of 2048 to build the USS system. On one hand, the segment prediction embedding achieves an SDR of 7.80 dB on AudioSet, outperforming the embedding condition of 5.57 dB. The segment prediction also achieves higher SDRis than the embedding condition on the FSDKaggle2018, and the FSD50k dataset datasets. An explaination is that the sound classes of all of the AudioSet, FSDKaggle2018, and the FSD50k datasets are sub-classes of the AudioSet. The segment prediction performs better than embedding condition in in-vocabulary sound classes separation. On the other hand, the embedding condition achieves higher SDRs than the segment prediction on the MUSDB18 and the Slakh2100 dataset. This result indicates that the embedding condition perform better than segment prediction in new-vocabulary sound classes separation.

Fig. \ref{fig:audioset_long_fig} in the end of this paper shows the classwise SDRi results of AudioSet separation including 527 sound classes. The dashed lines show the SDRi with oracle segment prediction or embedding as conditions. The solid lines show the SDRi with averaged segment prediction or embedding calculated from the anchor segments mined from the balanced training subset. Fig. \ref{fig:audioset_long_fig} shows that sound classes such as busy signal, sine wave, bicycle bell achieve the highest SDRi over 15 dB. We discovered that clear defined sound classes such as instruments can achieve high SDRi scores. Most of sound classes achieve positive SDRi scores. The tendency of using segment prediction and embedding as conditions are the same, although the segment prediction outperform the embedding and vice versa in some sound classes. 

\begin{table*}
  \caption{USS results with different sources number.}
  \label{table:mix_src_num}
  \centering
  \resizebox{\textwidth}{!}{
  \begin{tabular}{lccccccccccccccccccc}
    \toprule
    & \multicolumn{2}{c}{\textbf{AudioSet (SDRi)}} & \multicolumn{2}{c}{\textbf{FSDK2018}} & \multicolumn{2}{c}{\textbf{FSD50k}} & \multicolumn{4}{c}{\textbf{MUSDB18}} & \multicolumn{2}{c}{\textbf{Slakh2100}} & \multicolumn{2}{c}{\textbf{VoicebankDemand}} \\
	\cmidrule(lr){2-3} \cmidrule(lr){4-5} \cmidrule(lr){6-7}\cmidrule(lr){8-11}\cmidrule(lr){12-13}\cmidrule(lr){14-15}
    & ora. emb & avg. emb & SDR & SDRi & SDR & SDRi & SDR & SDRi & wSDR & wSDRi & SDR & SDRi & PESQ & SSNR  \\
    \midrule
    2 srcs to 1 src & 8.26 & 5.57 & 10.61 & 10.57 & 5.54 & 5.79 & 3.08 & 8.12 & 4.02 & 9.22 & -0.46 & 7.28 & 2.18 & 9.00 \\
    3 srcs to 1-2 srcs & 7.37 & 4.71 & 8.30 & 8.26 & 4.36 & 4.52 & 2.43 & 8.08 & 3.56 & 8.69 & -0.48 & 7.26 & 2.37 & 8.34 \\
    4 srcs to 1-3 srcs & 7.03 & 4.38 & 7.49 & 7.45 & 3.99 & 4.10 & 2.43 & 7.99 & 3.51 & 8.98 & 0.70 & 8.44 & 2.38 & 7.78 \\
    \bottomrule
\end{tabular}
}
\end{table*}

\begin{table*}
  \caption{USS results with different data augmentation.}
  \label{table:data_augmentation}
  \centering
  \resizebox{\textwidth}{!}{
  \begin{tabular}{lccccccccccccccccccc}
    \toprule
    & \multicolumn{2}{c}{\textbf{AudioSet (SDRi)}} & \multicolumn{2}{c}{\textbf{FSDK2018}} & \multicolumn{2}{c}{\textbf{FSD50k}} & \multicolumn{4}{c}{\textbf{MUSDB18}} & \multicolumn{2}{c}{\textbf{Slakh2100}} & \multicolumn{2}{c}{\textbf{VoicebankDemand}} \\
	\cmidrule(lr){2-3} \cmidrule(lr){4-5} \cmidrule(lr){6-7}\cmidrule(lr){8-11}\cmidrule(lr){12-13}\cmidrule(lr){14-15}
    & ora. emb & avg. emb & SDR & SDRi & SDR & SDRi & SDR & SDRi & wSDR & wSDRi & SDR & SDRi & PESQ & SSNR  \\
    \midrule
    no aug & 7.11 & 3.81 & 7.19 & 7.14 & 3.27 & 3.35 & 1.78 & 7.22 & 3.09 & 8.74 & 0.69 & 8.43 & 2.39 & 6.36 \\
    +- 20 dB & 5.51 & 3.62 & 5.77 & 5.73 & 2.93 & 2.94 & 1.69 & 7.02 & 2.51 & 8.03 & -0.34 & 7.40 & 2.22 & 5.34 \\
    +match energy & 8.26 & 5.57 & 10.61 & 10.57 & 5.54 & 5.79 & 3.08 & 8.12 & 4.02 & 9.22 & -0.46 & 7.28 & 2.18 & 9.00 \\
    \bottomrule
\end{tabular}
}
\end{table*}

\begin{table*}
  \caption{USS results of models trained with balanced and full subsets of AudioSet.}
  \label{table:balance_full}
  \centering
  \resizebox{\textwidth}{!}{
  \begin{tabular}{lccccccccccccccccccc}
    \toprule
    & \multicolumn{2}{c}{\textbf{AudioSet (SDRi)}} & \multicolumn{2}{c}{\textbf{FSDK2018}} & \multicolumn{2}{c}{\textbf{FSD50k}} & \multicolumn{4}{c}{\textbf{MUSDB18}} & \multicolumn{2}{c}{\textbf{Slakh2100}} & \multicolumn{2}{c}{\textbf{VoicebankDemand}} \\
	\cmidrule(lr){2-3} \cmidrule(lr){4-5} \cmidrule(lr){6-7}\cmidrule(lr){8-11}\cmidrule(lr){12-13}\cmidrule(lr){14-15}
    & ora. emb & avg. emb & SDR & SDRi & SDR & SDRi & SDR & SDRi & wSDR & wSDRi & SDR & SDRi & PESQ & SSNR  \\
    \midrule
    +Balanced set & 8.26 & 5.57 & 10.61 & 10.57 & 5.54 & 5.79 & 3.08 & 8.12 & 4.02 & 9.22 & -0.46 & 7.28 & 2.18 & 9.00 \\
    Full set & 8.21 & 6.14 & 10.34 & 10.30 & 5.45 & 5.71 & 2.31 & 7.20 & 3.60 & 8.92 & -1.29 & 6.45 & 2.40 & 9.45 \\
    Full set (long train) & 9.60 & 6.76 & 13.07 & 13.03 & 6.52 & 6.85 & 1.77 & 7.00 & 3.51 & 9.00 & -2.62 & 5.12 & 2.45 & 10.00 \\
    \bottomrule
\end{tabular}
}
\end{table*}

\subsubsection{Freeze, Finetune, and Adapt Conditional Embeddings}
Table \ref{tab:freeze_fintune_emb} shows the comparison of using random, frozen, finetuned, and adapted conditional embeddings to build the USS system. All the variations of the conditional embeddings extractors are based on the CNN14 architecture. Using random weights to extract conditional embeddings achieves an SDRi of 2.82 dB on AudioSet, compared to use pretrained CNN14 to extract conditional embeddings achieves an SDR 5.57 dB. We show that using random weights to extract conditional embeddings work for all USS tasks, such as achieves an SDRi of 5.91 dB on the FSDKaggle2018 dataset compared to the pretrained CNN14 embedding extractor of 10.57 dB.

Next, we experiment with learning the parameters of the conditional embedding extractor from scratch or finetune the weights from pretrained models. Table \ref{tab:freeze_fintune_emb} shows that neither the learning from scratch nor the finetuning approach improves the USS system performance. The learning from scratch approach and the finetuning approaches achieve SDRi of 2.41 dB and 3.38 dB on the FSDKaggle2018 dataset, even underperform the random weights of 5.91 dB. One possible explanation is that the parameters of the conditional embedding branch and the source separation branch are difficult to be jointly optimized when both of them are deep. The training falls to a collapse mode. Using the pretrained frozen CNN14 system as conditional embedding extractor significantly improves the SDRi to 10.57 dB on the FSDKaggle2018 dataset.

Based on the pretrained frozen CNN14 conditional embedding extractor, we propose to add a learnable shortcut or add an learnable adaptor on top of the CNN14 system. The learnable short cut has a CNN14 architecture with learnable parameters. Table \ref{tab:freeze_fintune_emb} shows that the learnable shortcut conditional embedding extractor achieves an SDR of 9.29 dB on FSDKaggle2018, less than using the pretrained frozen CNN14 conditional embedding extractor of 10.57 dB. One possible explanation is that the learnable shortcut destories the embedding information for source separation. The adaptor is a 2-layer fully connected neural network on top of the pretrained frozen CNN14 conditional embedding extractor. With the adaptor, we achieve an SDR of 11.10 dB and outperforms the CNN14 system. This result indicates that the adaptor is beneficial for the USS task.

\subsubsection{Separation Architectures}
Table \ref{table:models} shows the results of building USS systems with different source separation backbones. The open-unmix system \cite{stoter2019open} is a 3-layer bidirectional long short term memory (BLSTM) system. The BLSTM is applied on the mixture spectrogram to output the estimated clean spectrogram. The open-unmix system achieves an SDR of 3.39 dB on AudioSet separation and achieves a PESQ of 2.40 on the Voicebank-Demand speech enhancement task, indicating that the BLSTM backbone performs well for speech enhancement. The open-unmix system underperforms other backbone source separation systems in FSDKaggle2018, FSD50k, MUSDB18, and Slakh2100 separation, indicating that the capacity of the open-unmix system is not large enough to separate a wide range of sound classes. The ConvTasNet \cite{luo2019conv} is a time-domain source separation system consists of one-dimensional convolutional encoder and decoder layers. The ConvTasNet achieves an SDRi of 5.00 dB on AudioSet separation and outperforms the open-unmix system.

Our proposed UNet30 \cite{jansson2017singing} is an encoder-decoder convolutional architecture consists of 30 convolutional layers. The ResUNet30 \cite{kong2021decoupling} adds residual shortcuts in the encoder and decoder blocks in UNet30. The UNet30 and the ResUNet30 systesm achieve SDRis of 5.50 dB and 5.57 dB on AudioSet, outperforming the ConvTasNet by around 1 dB in all source separation tasks. We extend ResUNet30 to a deeper system ResUNet60 with 60 convolutiona layers. Table \ref{table:models} shows that ResUNet60 outperforms ResUNet30 by around 0.5 dB in all USS tasks. This result indicates that deeper architectures are beneficial for USS. 

\subsubsection{Different Anchor Segment Durations}\label{section:Anchor_segment_duration}
Table \ref{table:segment_seconds} shows the results of USS systems trained with different anchor segment durations ranging from 0.5 s to 10 s. The anchor segments are mined by a pretrained SED system as described in Section \ref{section:anchor}. On one hand, Table \ref{table:segment_seconds} shows that the separation scores increase with anchor segment durations increase from 0.5 s to 2 s and achieves the best SDRi of 5.57 dB at anchor segment of 2 s on AudioSet separation. This result shows that the anchor segment should be long enough to contain sufficient context information to build the USS system. On the other hand, the separation scores decrease with anchor segment durations decrease from 2 s to 10 s on all tasks. One possible explanation is that long anchor segment contain undesired interfering sounds that will impair the training of the USS system. Therefore, we use 2-second anchor segment in all other experiments.

\subsubsection{Different Anchor Segment Mining Strategies}
Table \ref{table:anchor_mining_strategy} shows the results of different anchor mining strategies. The in-clip random strategy randomly select two anchor segments from a same 10-second audio clip which significantly underperform the SED mining strategy in all of the source separation tasks. The out-clip random strategy randomly select two anchor segments from two different 10-second audio clips. The out-clip random strategy achieves an SDRi of 5.90 dB on AudioSet, outperforms the SED mining of 5.57 dB. On one hand, the out-clip random strategy also outperforms the SED mining strategy in FSDKaggle2018 and the FSD50k dataset. On the other hand, the SED mining strategy outperforms the out-clip random strategy in MUSDB18 and Slakh2100 source separation. Both the out-clip and the SED mining strategies outperform the in-clip random strategy.

\begin{figure*}[t]
  \centering
\centerline{\includegraphics[width=\textwidth]{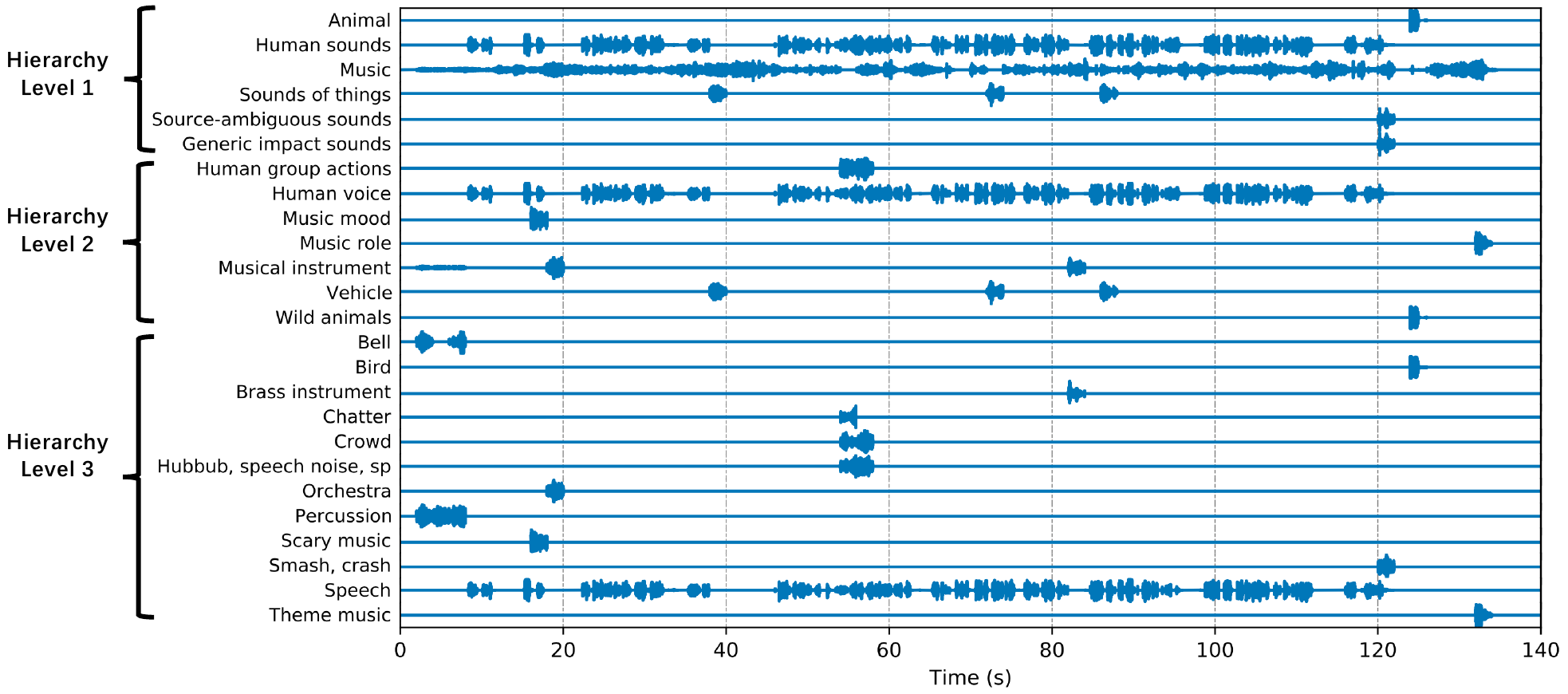}}
  \caption{Computational auditory scene analysis and hierarchical USS of the trailer of ``Harry Potter and the Sorcerer's Stone'': \url{https://www.youtube.com/watch?v=VyHV0BRtdxo}}
  \label{fig:harry_potter}
\end{figure*}

\subsubsection{Sources number to mix during training}
Table \ref{table:mix_src_num} shows the USS results trained with different number of sources $ J $ to constitute a mixture. Table \ref{table:mix_src_num} shows that $ J=2 $ performs the best on AudioSet with an SDRi of 5.57 dB and also performs the best on the FSDKaggle2018, FSD50k, and on MUSDB18 datasets. This result shows that mixing two sources is sufficient for those source separation tasks. By using $ J=4 $ the USS system perform the beston the Slakh2100 dataset. An explanation is that the Slakh2100 contains audio clips contain multiple instruments being played simultaneously. Using more sources to constitute a mixture perform better than using fewer sources.

\subsubsection{Data augmentation}\label{section:exp_data_augmentation}
Table \ref{table:data_augmentation} shows the USS results with different augmentation strategies applied to sources to create a mixture. First, we do not apply any data augmentation to create a mixture. Second, we randomly scale the volume of each source by $ \pm 20 $ dB. Third, we propose a matching energy data augmentation to scale the volume of sources to create a mixture to ensure the sources have the same energy. Table \ref{table:data_augmentation} shows that the matching energy data augmentation significantly outperform the systems trained without data augmentation and random volume scale augmentation, with an SDRi of 5.57 dB compared to 3.81 dB and 3.63 dB on AudioSet separation. The matching energy data augmentation also outperform no data augmentation and random volume augmentation on all the other tasks.

\subsubsection{USS results Trained with balanced and full AudioSet}
Table \ref{table:balance_full} shows the results of training the USS systems with the balanced and the full AudioSet, respectively. The full training data is 100 times larger than the balanced data. We also experiment with training the USS system with 4 GPUs and a larger batch size of 64. 
The USS system trained on the full AudioSet outperforms the USS system trained on the balanced set after trained 1 million steps. 
Table \ref{table:balance_full} shows that training on the full AudioSet with a batch size of 64 achieves an SDRi of 6.76 dB, outperforming training on the balanced set of 5.57 dB.

\subsubsection{Visualization of Hierarchical Separation}
One application of the hierarchical separation is to separate arbitrary audio recordings into individual sources with AudioSet ontology. For example, the USS system can separate the sound in a movie into different tracks. One challenge of the hierarchical separation is the number of present sources are unknown. We use the methods in Section \ref{section:hierarchy_separation} to detect and separate the present sound events.

Fig. \ref{fig:harry_potter} shows the automatically detected and separated waveforms of a movie clip from \textit{Harry Potter and the Sorcerer's Stone} from ontology levels 1 to 3 by using Algorithm \ref{alg:automatic_filt_sep}. Level 1 indicates coarse sound classes and level 3 indicates fine sound classes. In level 1, the USS system successfully separate human sounds, music and sounds of things. In level 2, the USS system further separate human group actions, vehicle, and animals. In level 3, the USS system separate fine-grained sound classes such as bell, bird, crowd, and scary music.

\section{Conclusion}
In this paper, we propose universal source separation (USS) systems trained on the large-scale weakly labelled AudioSet. The USS systems can separate hundreds of sound classes using a single model. The separation system can achieve universal source separation by using the embedding calculated from query examples as a condition. In training, we first apply a sound event detection (SED) system to detect the anchor segments that are most likely to contain sound events. We constitute a mixture by mixing several anchor segments. Then, we use a pretrained audio tagging system to calculate the segment prediction probability or the embedding vector as the condition of the target anchor segment. The USS system takes the mixture and the condition as input to output the desired anchor segment waveform. In inference, we propose both a hierarchical separation with an AudioSet ontology. We evaluated our proposed USS systems on a wide range of separation tasks, including AudioSet separation, FSDKaggle2018 and FSD50k general sound separation, MUSDB18 and Slakh2100 music instruments separation, and Voicebank-Demand speech enhancement without training on those datasets. We show the USS system is an approach that can address the USS problem. In future, we will improve the quality of the separated waveforms of the weakly labelled USS systems.

\bibliographystyle{IEEEtran}
\bibliography{refs}

\begin{thebibliography}{10}
\providecommand{\url}[1]{#1}
\csname url@samestyle\endcsname
\providecommand{\newblock}{\relax}
\providecommand{\bibinfo}[2]{#2}
\providecommand{\BIBentrySTDinterwordspacing}{\spaceskip=0pt\relax}
\providecommand{\BIBentryALTinterwordstretchfactor}{4}
\providecommand{\BIBentryALTinterwordspacing}{\spaceskip=\fontdimen2\font plus
\BIBentryALTinterwordstretchfactor\fontdimen3\font minus
  \fontdimen4\font\relax}
\providecommand{\BIBforeignlanguage}[2]{{%
\expandafter\ifx\csname l@#1\endcsname\relax
\typeout{** WARNING: IEEEtran.bst: No hyphenation pattern has been}%
\typeout{** loaded for the language `#1'. Using the pattern for}%
\typeout{** the default language instead.}%
\else
\language=\csname l@#1\endcsname
\fi
#2}}
\providecommand{\BIBdecl}{\relax}
\BIBdecl

\bibitem{loizou2007speech}
P.~C. Loizou, \emph{{Speech Enhancement: Theory and Practice}}.\hskip 1em plus
  0.5em minus 0.4em\relax CRC press, 2007.

\bibitem{xu2014regression}
Y.~Xu, J.~Du, L.-R. Dai, and C.-H. Lee, ``A regression approach to speech
  enhancement based on deep neural networks,'' \emph{IEEE/ACM Transactions on
  Audio, Speech, and Language Processing}, vol.~23, no.~1, pp. 7--19, 2014.

\bibitem{stoter20182018}
F.-R. St{\"o}ter, A.~Liutkus, and N.~Ito, ``The 2018 signal separation
  evaluation campaign,'' in \emph{International Conference on Latent Variable
  Analysis and Signal Separation}.\hskip 1em plus 0.5em minus 0.4em\relax
  Springer, 2018, pp. 293--305.

\bibitem{heittola2011sound}
T.~Heittola, A.~Mesaros, T.~Virtanen, and A.~Eronen, ``Sound event detection in
  multisource environments using source separation,'' in \emph{CHiME Workshop
  on Machine Listening in Multisource Environments (CHiME 2011)}, 2011, pp.
  36--40.

\bibitem{turpault2020improving}
N.~Turpault, S.~Wisdom, H.~Erdogan, J.~Hershey, R.~Serizel, E.~Fonseca,
  P.~Seetharaman, and J.~Salamon, ``Improving sound event detection in domestic
  environments using sound separation,'' in \emph{Detection and Classification
  of Acoustic Scenes and Events (DCASE) Workshop}, 2020.

\bibitem{haykin2005cocktail}
S.~Haykin and Z.~Chen, ``The cocktail party problem,'' \emph{Neural
  Computation}, vol.~17, no.~9, pp. 1875--1902, 2005.

\bibitem{brown1994computational}
G.~J. Brown and M.~Cooke, ``{Computational Auditory Scene Analysis},''
  \emph{Computer Speech \& Language}, vol.~8, no.~4, pp. 297--336, 1994.

\bibitem{wang2006computational}
D.~Wang and G.~J. Brown, \emph{Computational auditory scene analysis:
  Principles, algorithms, and applications}.\hskip 1em plus 0.5em minus
  0.4em\relax Wiley-IEEE press, 2006.

\bibitem{stoller2018wave}
D.~Stoller, S.~Ewert, and S.~Dixon, ``{Wave-U-Net: A multi-scale neural network
  for end-to-end audio source separation},'' in \emph{International Society for
  Music Information Retrieval (ISMIR)}, 2018.

\bibitem{luo2021rethinking}
Y.~Luo, Z.~Chen, C.~Han, C.~Li, T.~Zhou, and N.~Mesgarani, ``Rethinking the
  separation layers in speech separation networks,'' in \emph{IEEE
  International Conference on Acoustics, Speech and Signal Processing
  (ICASSP)}.\hskip 1em plus 0.5em minus 0.4em\relax IEEE, 2021, pp. 1--5.

\bibitem{kavalerov2019universal}
I.~Kavalerov, S.~Wisdom, H.~Erdogan, B.~Patton, K.~Wilson, J.~Le~Roux, and
  J.~R. Hershey, ``Universal sound separation,'' in \emph{IEEE Workshop on
  Applications of Signal Processing to Audio and Acoustics (WASPAA)}, 2019, pp.
  175--179.

\bibitem{wisdom2020unsupervised}
S.~Wisdom, E.~Tzinis, H.~Erdogan, R.~J. Weiss, K.~Wilson, and J.~R. Hershey,
  ``Unsupervised speech separation using mixtures of mixtures,'' in
  \emph{Neural Information Processing Systems (NeurIPS)}, 2020.

\bibitem{tzinis2020sudo}
E.~Tzinis, Z.~Wang, and P.~Smaragdis, ``Sudo rm-rf: Efficient networks for
  universal audio source separation,'' in \emph{IEEE International Workshop on
  Machine Learning for Signal Processing (MLSP)}, 2020.

\bibitem{seetharaman2019class}
P.~Seetharaman, G.~Wichern, S.~Venkataramani, and J.~Le~Roux,
  ``Class-conditional embeddings for music source separation,'' in \emph{IEEE
  International Conference on Acoustics, Speech and Signal Processing
  (ICASSP)}, 2019, pp. 301--305.

\bibitem{tzinis2020improving}
E.~Tzinis, S.~Wisdom, J.~R. Hershey, A.~Jansen, and D.~P. Ellis, ``Improving
  universal sound separation using sound classification,'' in \emph{IEEE
  International Conference on Acoustics, Speech and Signal Processing
  (ICASSP)}, 2020, pp. 96--100.

\bibitem{wang2018voicefilter}
Q.~Wang, H.~Muckenhirn, K.~Wilson, P.~Sridhar, Z.~Wu, J.~Hershey, R.~A.
  Saurous, R.~J. Weiss, Y.~Jia, and I.~L. Moreno, ``Voicefilter: Targeted voice
  separation by speaker-conditioned spectrogram masking,'' in
  \emph{INTERSPEECH}, 2019.

\bibitem{gfeller2021one}
B.~Gfeller, D.~Roblek, and M.~Tagliasacchi, ``One-shot conditional audio
  filtering of arbitrary sounds,'' in \emph{IEEE International Conference on
  Acoustics, Speech and Signal Processing (ICASSP)}, 2021, pp. 501--505.

\bibitem{choi2021lasaft}
W.~Choi, M.~Kim, J.~Chung, and S.~Jung, ``{LaSAFT: Latent Source Attentive
  Frequency Transformation for Conditioned Source Separation},'' in \emph{IEEE
  International Conference on Acoustics, Speech and Signal Processing
  (ICASSP)}, 2021, pp. 171--175.

\bibitem{pishdadian2020finding}
F.~Pishdadian, G.~Wichern, and J.~Le~Roux, ``Learning to separate sounds from
  weakly labeled scenes,'' in \emph{IEEE International Conference on Acoustics,
  Speech and Signal Processing (ICASSP)}, 2020, pp. 91--95.

\bibitem{tzinis2021compute}
E.~Tzinis, Z.~Wang, X.~Jiang, and P.~Smaragdis, ``Compute and memory efficient
  universal sound source separation,'' \emph{Journal of Signal Processing
  Systems}, pp. 245–--259, 2021.

\bibitem{liu2022separate}
X.~Liu, H.~Liu, Q.~Kong, X.~Mei, J.~Zhao, Q.~Huang, M.~D. Plumbley, and
  W.~Wang, ``Separate what you describe: Language-queried audio source
  separation,'' in \emph{INTERSPEECH}, 2022.

\bibitem{mesaros2019sound}
A.~Mesaros, A.~Diment, B.~Elizalde, T.~Heittola, E.~Vincent, B.~Raj, and
  T.~Virtanen, ``{Sound event detection in the DCASE 2017 challenge},''
  \emph{IEEE/ACM Transactions on Audio, Speech, and Language Processing},
  vol.~27, no.~6, pp. 992--1006, 2019.

\bibitem{gemmeke2017audio}
J.~F. Gemmeke, D.~P. Ellis, D.~Freedman, A.~Jansen, W.~Lawrence, R.~C. Moore,
  M.~Plakal, and M.~Ritter, ``Audio set: An ontology and human-labeled dataset
  for audio events,'' in \emph{IEEE International Conference on Acoustics,
  Speech and Signal Processing (ICASSP)}, 2017, pp. 776--780.

\bibitem{shah2018closer}
A.~Shah, A.~Kumar, A.~G. Hauptmann, and B.~Raj, ``A closer look at weak label
  learning for audio events,'' \emph{arXiv preprint arXiv:1804.09288}, 2018.

\bibitem{kong2020panns}
Q.~Kong, Y.~Cao, T.~Iqbal, Y.~Wang, W.~Wang, and M.~D. Plumbley, ``{PANNs:
  Large-scale pretrained audio neural networks for audio pattern
  recognition},'' \emph{IEEE/ACM Transactions on Audio, Speech, and Language
  Processing}, vol.~28, pp. 2880--2894, 2020.

\bibitem{gong2021psla}
Y.~Gong, Y.-A. Chung, and J.~Glass, ``{PSLA: Improving audio tagging with
  pretraining, sampling, labeling, and aggregation},'' \emph{IEEE/ACM
  Transactions on Audio, Speech, and Language Processing}, vol.~29, pp.
  3292--3306, 2021.

\bibitem{chen2022hts}
K.~Chen, X.~Du, B.~Zhu, Z.~Ma, T.~Berg-Kirkpatrick, and S.~Dubnov, ``{HTS-AT: A
  hierarchical token-semantic audio transformer for sound classification and
  detection},'' in \emph{IEEE International Conference on Acoustics, Speech and
  Signal Processing (ICASSP)}, 2022, pp. 646--650.

\bibitem{kong2017joint}
Q.~Kong, Y.~Xu, W.~Wang, and M.~D. Plumbley, ``A joint detection-classification
  model for audio tagging of weakly labelled data,'' in \emph{IEEE
  International Conference on Acoustics, Speech and Signal Processing
  (ICASSP)}, 2017, pp. 641--645.

\bibitem{wang2018polyphonic}
Y.~Wang, ``Polyphonic sound event detection with weak labeling,'' \emph{PhD
  thesis}, 2018.

\bibitem{adavanne2019sound}
S.~Adavanne, H.~Fayek, and V.~Tourbabin, ``Sound event classification and
  detection with weakly labeled data,'' in \emph{Detection and Classification
  of Acoustic Scenes and Events (DCASE) Workshop}, 2019.

\bibitem{kong2020source}
Q.~Kong, Y.~Wang, X.~Song, Y.~Cao, W.~Wang, and M.~D. Plumbley, ``{Source
  separation with weakly labelled data: An approach to computational auditory
  scene analysis},'' in \emph{IEEE International Conference on Acoustics,
  Speech and Signal Processing (ICASSP)}, 2020, pp. 101--105.

\bibitem{chen2022zero}
K.~Chen, X.~Du, B.~Zhu, Z.~Ma, T.~Berg-Kirkpatrick, and S.~Dubnov, ``Zero-shot
  audio source separation through query-based learning from weakly-labeled
  data,'' in \emph{Proceedings of the AAAI Conference on Artificial
  Intelligence}, vol.~36, no.~4, 2022, pp. 4441--4449.

\bibitem{kong2021decoupling}
Q.~Kong, Y.~Cao, H.~Liu, K.~Choi, and Y.~Wang, ``Decoupling magnitude and phase
  estimation with deep resunet for music source separation,'' in
  \emph{International Society for Music Information Retrieval (ISMIR)}, 2021.

\bibitem{nmf}
D.~D. Lee and H.~S. Seung, ``Algorithms for non-negative matrix
  factorization,'' in \emph{Neural Information Processing Systems (NeurIPS)},
  2000.

\bibitem{defossez2019demucs}
A.~D{\'e}fossez, N.~Usunier, L.~Bottou, and F.~Bach, ``Demucs: Deep extractor
  for music sources with extra unlabeled data remixed,'' \emph{arXiv preprint
  arXiv:1909.01174}, 2019.

\bibitem{defossez2019music}
------, ``Music source separation in the waveform domain,'' \emph{arXiv
  preprint arXiv:1911.13254}, 2019.

\bibitem{luo2018tasnet}
Y.~Luo and N.~Mesgarani, ``Tasnet: time-domain audio separation network for
  real-time, single-channel speech separation,'' in \emph{IEEE International
  Conference on Acoustics, Speech and Signal Processing (ICASSP)}, 2018, pp.
  696--700.

\bibitem{luo2019conv}
------, ``{Conv-TasNet: Surpassing ideal time--frequency magnitude masking for
  speech separation},'' \emph{IEEE/ACM transactions on audio, speech, and
  language processing}, vol.~27, no.~8, pp. 1256--1266, 2019.

\bibitem{luo2022music}
Y.~Luo and J.~Yu, ``Music source separation with band-split rnn,'' \emph{arXiv
  preprint arXiv:2209.15174}, 2022.

\bibitem{narayanan2013ideal}
A.~Narayanan and D.~Wang, ``Ideal ratio mask estimation using deep neural
  networks for robust speech recognition,'' in \emph{IEEE International
  Conference on Acoustics, Speech and Signal Processing}, 2013, pp. 7092--7096.

\bibitem{williamson2015complex}
D.~S. Williamson, Y.~Wang, and D.~Wang, ``Complex ratio masking for monaural
  speech separation,'' \emph{IEEE/ACM Transactions on Audio, Speech, and
  Language Processing}, vol.~24, no.~3, pp. 483--492, 2015.

\bibitem{huang2015joint}
P.-S. Huang, M.~Kim, M.~Hasegawa-Johnson, and P.~Smaragdis, ``Joint
  optimization of masks and deep recurrent neural networks for monaural source
  separation,'' \emph{IEEE/ACM Transactions on Audio, Speech, and Language
  Processing}, vol.~23, no.~12, pp. 2136--2147, 2015.

\bibitem{takahashi2018mmdenselstm}
N.~Takahashi, N.~Goswami, and Y.~Mitsufuji, ``{MMDenseLSTM: An efficient
  combination of convolutional and recurrent neural networks for audio source
  separation},'' in \emph{IEEE International Workshop on Acoustic Signal
  Enhancement (IWAENC)}, 2018, pp. 106--110.

\bibitem{uhlich2017improving}
S.~Uhlich, M.~Porcu, F.~Giron, M.~Enenkl, T.~Kemp, N.~Takahashi, and
  Y.~Mitsufuji, ``Improving music source separation based on deep neural
  networks through data augmentation and network blending,'' in \emph{IEEE
  International Conference on Acoustics, Speech and Signal Processing
  (ICASSP)}, 2017, pp. 261--265.

\bibitem{chandna2017monoaural}
P.~Chandna, M.~Miron, J.~Janer, and E.~G{\'o}mez, ``Monoaural audio source
  separation using deep convolutional neural networks,'' in \emph{International
  Conference on Latent Variable Analysis and Signal Separation}.\hskip 1em plus
  0.5em minus 0.4em\relax Springer, 2017, pp. 258--266.

\bibitem{hu2020dccrn}
Y.~Hu, Y.~Liu, S.~Lv, M.~Xing, S.~Zhang, Y.~Fu, J.~Wu, B.~Zhang, and L.~Xie,
  ``{DCCRN: Deep complex convolution recurrent network for phase-aware speech
  enhancement},'' in \emph{INTERSPEECH}, 2020.

\bibitem{jansson2017singing}
A.~Jansson, E.~Humphrey, N.~Montecchio, R.~Bittner, A.~Kumar, and T.~Weyde,
  ``{Singing voice separation with deep U-Net convolutional networks},'' in
  \emph{International Society for Music Information Retrieval (ISMIR)}, 2017.

\bibitem{hennequin2020spleeter}
R.~Hennequin, A.~Khlif, F.~Voituret, and M.~Moussallam, ``Spleeter: a fast and
  efficient music source separation tool with pre-trained models,''
  \emph{Journal of Open Source Software}, vol.~5, no.~50, p. 2154, 2020.

\bibitem{defossez2021hybrid}
A.~D{\'e}fossez, ``Hybrid spectrogram and waveform source separation,'' in
  \emph{Proceedings of the ISMIR 2021 Workshop on Music Source Separation},
  2021.

\bibitem{rouard2022hybrid}
S.~Rouard, F.~Massa, and A.~D{\'e}fossez, ``Hybrid transformers for music
  source separation,'' in \emph{International Conference on Acoustics, Speech,
  and Signal Processing (ICASSP)}, 2023.

\bibitem{veaux2013voice}
C.~Veaux, J.~Yamagishi, and S.~King, ``{The Voice Bank Corpus: Design,
  collection and data analysis of a large regional accent speech database},''
  in \emph{International Conference Oriental COCOSDA with Conference on Asian
  Spoken Language Research and Evaluation (O-COCOSDA/CASLRE)}, 2013.

\bibitem{rafii2017musdb18}
\BIBentryALTinterwordspacing
Z.~Rafii, A.~Liutkus, F.-R. St{\"o}ter, S.~I. Mimilakis, and R.~Bittner, ``The
  {MUSDB18} corpus for music separation,'' Dec. 2017. [Online]. Available:
  \url{https://doi.org/10.5281/zenodo.1117372}
\BIBentrySTDinterwordspacing

\bibitem{salamon2014dataset}
J.~Salamon, C.~Jacoby, and J.~P. Bello, ``A dataset and taxonomy for urban
  sound research,'' in \emph{Proceedings of the 22nd ACM international
  conference on Multimedia}, 2014, pp. 1041--1044.

\bibitem{fonseca2018general}
E.~Fonseca, M.~Plakal, F.~Font, D.~P. Ellis, X.~Favory, J.~Pons, and X.~Serra,
  ``{General-purpose tagging of Freesound audio with Audioset labels: Task
  description, dataset, and baseline},'' in \emph{Proceedings of the Detection
  and Classification of Acoustic Scenes and Events 2018 Workshop (DCASE)},
  2018.

\bibitem{wisdom2021s}
S.~Wisdom, H.~Erdogan, D.~P. Ellis, R.~Serizel, N.~Turpault, E.~Fonseca,
  J.~Salamon, P.~Seetharaman, and J.~R. Hershey, ``What’s all the fuss about
  free universal sound separation data?'' in \emph{IEEE International
  Conference on Acoustics, Speech and Signal Processing (ICASSP)}, 2021, pp.
  186--190.

\bibitem{gong2021ast}
Y.~Gong, Y.-A. Chung, and J.~Glass, ``{AST: Audio spectrogram transformer},''
  in \emph{INTERSPEECH}, 2021.

\bibitem{hershey2017cnn}
S.~Hershey, S.~Chaudhuri, D.~P. Ellis, J.~F. Gemmeke, A.~Jansen, R.~C. Moore,
  M.~Plakal, D.~Platt, R.~A. Saurous, B.~Seybold \emph{et~al.}, ``{CNN
  architectures for large-scale audio classification},'' in \emph{IEEE
  International Conference on Acoustics, Speech and Signal Processing
  (ICASSP)}, 2017, pp. 131--135.

\bibitem{wang2019comparison}
Y.~Wang, J.~Li, and F.~Metze, ``A comparison of five multiple instance learning
  pooling functions for sound event detection with weak labeling,'' in
  \emph{IEEE International Conference on Acoustics, Speech and Signal
  Processing (ICASSP)}, 2019, pp. 31--35.

\bibitem{liu2021swin}
Z.~Liu, Y.~Lin, Y.~Cao, H.~Hu, Y.~Wei, Z.~Zhang, S.~Lin, and B.~Guo, ``{Swin
  transformer: Hierarchical vision transformer using shifted windows},'' in
  \emph{Proceedings of the IEEE/CVF International Conference on Computer Vision
  (ICCV)}, 2021, pp. 10\,012--10\,022.

\bibitem{delcroix2022soundbeam}
M.~Delcroix, J.~B. V{\'a}zquez, T.~Ochiai, K.~Kinoshita, Y.~Ohishi, and
  S.~Araki, ``{SoundBeam: Target sound extraction conditioned on sound-class
  labels and enrollment clues for increased performance and continuous
  learning},'' \emph{IEEE/ACM Transactions on Audio, Speech, and Language
  Processing}, 2022.

\bibitem{meseguer2019conditioned}
G.~Meseguer-Brocal and G.~Peeters, ``{Conditioned-U-Net: Introducing a control
  mechanism in the U-Net for multiple source separations},'' \emph{arXiv
  preprint arXiv:1907.01277}, 2019.

\bibitem{ioffe2015batch}
S.~Ioffe and C.~Szegedy, ``{Batch normalization: Accelerating deep network
  training by reducing internal covariate shift},'' in \emph{International
  Conference on Machine Learning}, 2015, pp. 448--456.

\bibitem{perez2018film}
E.~Perez, F.~Strub, H.~De~Vries, V.~Dumoulin, and A.~Courville, ``{FiLM: Visual
  reasoning with a general conditioning layer},'' in \emph{Proceedings of the
  AAAI Conference on Artificial Intelligence}, vol.~32, no.~1, 2018.

\bibitem{he2016identity}
K.~He, X.~Zhang, S.~Ren, and J.~Sun, ``Identity mappings in deep residual
  networks,'' in \emph{European Conference on Computer Vision}.\hskip 1em plus
  0.5em minus 0.4em\relax Springer, 2016, pp. 630--645.

\bibitem{fonseca2020fsd50k}
E.~Fonseca, X.~Favory, J.~Pons, F.~Font, and X.~Serra, ``{FSD50k: an open
  dataset of human-labeled sound events},'' \emph{IEEE/ACM Transactions on
  Audio, Speech, and Language Processing}, 2020.

\bibitem{manilow2019cutting}
E.~Manilow, G.~Wichern, P.~Seetharaman, and J.~Le~Roux, ``Cutting music source
  separation some {Slakh}: A dataset to study the impact of training data
  quality and quantity,'' in \emph{IEEE Workshop on Applications of Signal
  Processing to Audio and Acoustics (WASPAA)}, 2019.

\bibitem{stoter2019open}
F.-R. St{\"o}ter, S.~Uhlich, A.~Liutkus, and Y.~Mitsufuji, ``{Open-Unmix - A
  reference implementation for music source separation},'' \emph{Journal of
  Open Source Software}, vol.~4, no.~41, p. 1667, 2019.

\bibitem{xu2015empirical}
B.~Xu, N.~Wang, T.~Chen, and M.~Li, ``Empirical evaluation of rectified
  activations in convolutional network,'' \emph{arXiv preprint
  arXiv:1505.00853}, 2015.

\bibitem{kingma2014adam}
D.~P. Kingma and J.~Ba, ``Adam: A method for stochastic optimization,'' in
  \emph{International Conference on Learning Representations (ICLR)}, 2015.

\bibitem{thiemann2013demand}
J.~Thiemann, N.~Ito, and E.~Vincent, ``{DEMAND: a collection of multi-channel
  recordings of acoustic noise in diverse environments},'' in \emph{Proceedings
  of Meetings on Acoustics}, 2013.

\bibitem{vincent2006performance}
E.~Vincent, R.~Gribonval, and C.~F{\'e}votte, ``Performance measurement in
  blind audio source separation,'' \emph{IEEE Transactions on Audio, Speech,
  and Language Processing}, vol.~14, no.~4, pp. 1462--1469, 2006.

\bibitem{recommendation2001perceptual}
ITU-T, ``{Perceptual evaluation of speech quality (PESQ): An objective method
  for end-to-end speech quality assessment of narrow-band telephone networks
  and speech codecs},'' \emph{Rec. ITU-T P. 862}, 2001.

\bibitem{quackenbush1988objective}
S.~R. Quackenbush, T.~P. Barnwell, and M.~A. Clements, \emph{Objective Measures
  of Speech Quality}.\hskip 1em plus 0.5em minus 0.4em\relax Prentice Hall,
  1988.

\bibitem{schneider2019wav2vec}
S.~Schneider, A.~Baevski, R.~Collobert, and M.~Auli, ``wav2vec: Unsupervised
  pre-training for speech recognition,'' in \emph{INTERSPEECH}, 2019.

\end{thebibliography}

\clearpage
\onecolumn
\appendix

\begin{figure*}[h]
  \centering
  \includegraphics[width=1.0\textwidth]{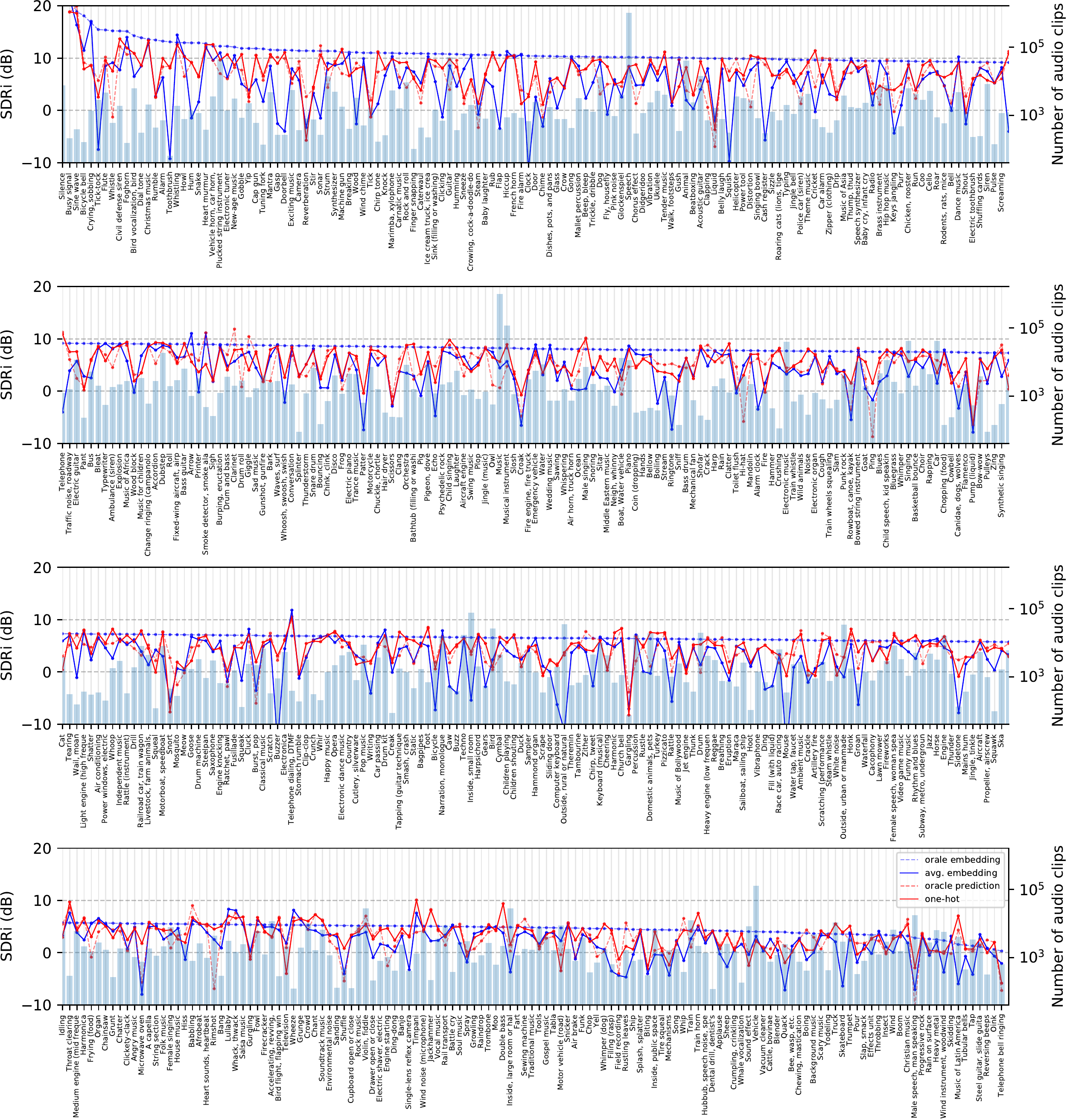}
  \caption{USS result on 527 AudioSet sound classes.}
  \label{fig:audioset_long_fig}
\end{figure*}


\end{document}